\documentclass[12pt,amsmath,amssymb,showkeys]{revtex4}
\usepackage{graphicx}
\usepackage{bm}
\usepackage{bbold}
\usepackage{hhline}

\def\br{ \bm{r} }
\def\bk{ \bm{k} }

\def\hbp{ \hat{\bm{p}} }

\def\bgam{ \bm{\gamma} }

\def\bB{ \bm{B} }

\begin{document}
\title{On the pseudospin description of the electron Bloch bands}

\author{K. V. Samokhin\footnote{E-mail: kirill.samokhin@brocku.ca}}
\affiliation{Department of Physics, Brock University, St. Catharines, Ontario L2S 3A1, Canada}


\begin{abstract}
We study the transformation properties of the electron states in crystals with spin-orbit coupling, focusing primarily on the limitations of the frequently used pseudospin-$1/2$ description of twofold degenerate Bloch
bands. Using the language of corepresentations of magnetic point groups, we construct the Bloch bases across the Brillouin zone in a way which is consistent with all symmetry requirements. This construction is applied to
derive the effective spin-orbit Hamiltonians in noncentrosymmetric crystals, known as the generalized Rashba models, in both single-band and multiband cases. 
\end{abstract}


\maketitle

\newpage
\section{Introduction}
\label{sec: Intro}

A textbook result of the quantum theory of solids is that the electron bands in a crystal which has both time reversal (TR) and inversion symmetries are at least twofold degenerate at each wave vector $\bk$ in the 
first Brillouin zone (BZ) \cite{Kittel-book}. The reason is that the Bloch states $|\bk\rangle$ and $KI|\bk\rangle$ belong to the same $\bk$ and are orthogonal. Here $K$ is the TR operation acting on spin-1/2 wave functions and 
$I$ is the space inversion operation. These two degenerate states are labelled by the index $s=1,2$. 

Due to the inevitable presence of the electron-lattice spin-orbit spin-orbit coupling (SOC), the states $|\bk,1\rangle$ and $|\bk,2\rangle=KI|\bk,1\rangle$ are not pure spin eigenstates. 
It is usually assumed that these states can nevertheless be chosen to have the same transformation properties under the crystal point group operations and 
TR as the pure spin-1/2 states, hence the name ``pseudospin'' for $s$. Then, the orientations of the pseudospin Bloch bases at different $\bk$ are defined by the Ueda-Rice formula \cite{UR85} 
(more recent discussions of different ways to construct the pseudospin bases across the whole BZ can be found in Refs. \cite{Yip13} and \cite{Fu15}). The Ueda-Rice construction has been extensively used in various applications. 
It forms, for instance, the foundation of the symmetry-based approach to the classification of unconventional superconducting states, 
see Refs. \cite{And84,VG85,UR85,Blount85,SU-review,TheBook}. However, the universal applicability of the pseudospin-$1/2$ picture has been questioned recently, in particular, in the context of the ``$j=3/2$'' pairing 
\cite{Brydon16,VSRLF18,Kim18} and in multiorbital systems \cite{NHI16}. 
Also, it has been shown that the standard superconducting gap symmetry classification can break down in nonsymmorphic crystals \cite{YM92,Norman95,MN17}. 

The goal of this article is to systematically analyze the symmetry properties of the electron Bloch states in TR-invariant crystals, both with and without inversion symmetry, in the presence of an arbitrarily strong 
SOC, focusing, in particular, on the validity of a pseudospin-$1/2$ description and possible reasons for its failure. We will show how to modify the Ueda-Rice formula in the non-pseudospin cases 
and consistently define the Bloch bases across the BZ in any twofold degenerate band. Due to the crucial role played by the TR symmetry, which is described by an antiunitary operator, we find it convenient 
to use the language of corepresentations of magnetic point groups, instead of the usual group representations. We will derive the effective model Hamiltonians of the electron-lattice SOC in crystals without an inversion center, 
in both pseudospin and non-pseudospin cases. Such models have numerous applications in many contexts, in particular, in the burgeoning field of topological materials \cite{Bernevig-Book}.

The article is organized as follows. In Secs. \ref{sec: band structure} and \ref{sec: G_k} we introduce the corepresentations terminology and notations and discuss the ``local'' symmetry properties of the Bloch states in
the reciprocal space. In Sec. \ref{sec: whole-BZ}, we show how to construct a ``global'' Bloch basis in the whole BZ which is compatible with all local symmetry requirements. In Sec. \ref{sec: ASOC},
the generalized Rashba Hamiltonians are derived, in the single-band and two-band cases. Sec. \ref{sec: Conclusion} concludes with a discussion of our results.

\section{Bloch states in the presence of spin-orbit coupling}
\label{sec: band structure}

Our starting point is the following Hamiltonian for non-interacting electrons in a crystal:
\begin{equation}
\label{general H}
    \hat H=\frac{\hbp^2}{2m}+U(\br)
    +\frac{\hbar}{4m^2c^2}\hat{\bm{\sigma}}[\bm{\nabla}U(\br)\times\hbp],
\end{equation}
where $\hbp=-i\hbar\bm{\nabla}$ is the momentum operator, $U(\br)$ is the lattice potential, and $\hat{\bm{\sigma}}=(\hat\sigma_1,\hat\sigma_2,\hat\sigma_3)$ are the
Pauli matrices. The last term is the electron-lattice SOC, which is not assumed to be small. We neglect impurities, lattice defects, and phonons, so that the Hamiltonian has the perfect periodicity of a Bravais lattice. 
In this section, as well as in Secs. \ref{sec: G_k} and \ref{sec: whole-BZ} below, we assume that the crystal has an inversion center, therefore $U(-\br)=U(\br)$. 

The symmetry operations leaving the crystal lattice invariant form the space group of the crystal. We consider only symmorphic space groups, which are generated by the Bravais lattice translations and the crystallographic 
point group operations (rotations, reflections, and inversion $I$). The point group is denoted by $\mathbb{G}$. In addition to the space group operations, the Hamiltonian (\ref{general H}) is also invariant under time reversal $K$. 

The eigenstates of the Hamiltonian (\ref{general H}) are given by the spinor Bloch functions $|\bk,n,s\rangle$, labelled by the wave vector $\bk$, which takes values in the BZ, and by the band index $n$. 
The corresponding eigenvalues form the bands $\epsilon_n(\bk)$, which are at least twofold degenerate at each $\bk$ due to the combined symmetry operation ${\cal C}=KI$, called ``conjugation'' \cite{Kittel-book}. 
The additional index $s=1,2$ distinguishes two orthonormal states within the same band: 
\begin{equation}
\label{Bloch pseudospinors}
  |\bk,n,1\rangle,\quad |\bk,n,2\rangle\equiv {\cal C}|\bk,n,1\rangle,
\end{equation}
or, explicitly:
\begin{equation}
\label{Bloch-states-1-2}
  |\bk,n,1\rangle=\frac{1}{\sqrt{\cal V}}
    \begin{pmatrix} u_{\bk,n}(\br)\\ v_{\bk,n}(\br)\end{pmatrix} e^{i\bk\br},\quad 
  |\bk,n,2\rangle=\frac{1}{\sqrt{\cal V}}
    \begin{pmatrix} -v^*_{\bk,n}(-\br)\\ u^*_{\bk,n}(-\br) \end{pmatrix} e^{i\bk\br},
\end{equation}
where ${\cal V}$ is the system volume and the Bloch factors $u_{\bk,n}(\br)$ and $v_{\bk,n}(\br)$ have the same periodicity as the crystal lattice. Note that ${\cal C}|\bk,n,2\rangle=-|\bk,n,1\rangle$, therefore
${\cal C}^2=-1$. The four states $|\pm\bk,n,1\rangle$, $|\pm\bk,n,2\rangle$ have the same energy $\epsilon_n(\bk)=\epsilon_n(-\bk)$.

We can drop the band index $n$ for brevity and ask the following question: Do the conjugate Bloch states $|\bk,1\rangle$ and $|\bk,2\rangle$ form a pseudospin-1/2 basis? In other words, do they 
transform under the point group operations in the same way as the pure spin-1/2 states, or the basis spinors, 
$\xi_1\equiv\xi_\uparrow$ and $\xi_2\equiv\xi_\downarrow$? We recall that the basis spinors transform under 
a proper rotation $R$ through an angle $\theta$ about an axis $\bm{n}$ as follows:
\begin{equation}
\label{xis-R}
  R(\bm{n},\theta)\xi_s=\sum_{s'}\xi_{s'}D^{(1/2)}_{s's}(R),
\end{equation}
where 
\begin{equation}
\label{D-1/2}
  \hat D^{(1/2)}(R)=e^{-i\theta(\bm{n}\hat{\bm{\sigma}})/2}
\end{equation}
is the spin-1/2 representation of rotations, see, for instance, Ref. \cite{Lax-book}. The basis spinors are not affected by inversion, $I\xi_s=\xi_s$, and, 
since a mirror reflection in a plane can be represented as a product of inversion and a $\pi$ rotation about the normal vector to the plane, \textit{i.e.},  
$\sigma_{\bm{n}}=IC_{2\bm{n}}$, we have:
\begin{equation}
\label{xis-sigma}
  \sigma_{\bm{n}}\xi_s=-i\sum_{s'}\xi_{s'}(\bm{n}\hat{\bm{\sigma}})_{s's}.
\end{equation}
The transformation under TR and conjugation is given by
\begin{equation}
\label{xis-TR-C}
  K(c\xi_1)=c^*\xi_2,\quad K(c\xi_2)=-c^*\xi_1,\qquad {\cal C}(c\xi_1)=c^*\xi_2,\quad {\cal C}(c\xi_2)=-c^*\xi_1,
\end{equation}
where we included a $c$-number coefficient to emphasize the antilinearity of the $K$ and ${\cal C}$ operators. 

The fact that the Bloch states $|\bk,1\rangle$ and $|\bk,2\rangle$ depend on the wave vector $\bk$, which is itself affected by the point group operations, means that the pseudospin property should be established 
separately for the operations leaving $\bk$ invariant and for those changing $\bk$. In the former case, discussed in Sec. \ref{sec: G_k}, one can work locally in the reciprocal space by analyzing the symmetry 
of the states with a given wave vector $\bk$. In the latter case, see Sec. \ref{sec: whole-BZ}, the point group operations take $\bk$ into a different ray of the star of $\bk$, so that 
the choice of the relative orientation of the Bloch bases at different points in the BZ becomes important. 

The point group operations that leave a given wave vector $\bk$ unchanged form a subgroup of $\mathbb{G}$, which we denote by $G_{\bk}$ and call the group of $\bk$ (in the literature, this group is also called
the little co-group of $\bk$ and denoted by $\bar G^{\bk}$, see Ref. \cite{BC-book}). We note that, while the invariance of $\bk$ should, in general, be taken modulo a reciprocal lattice vector $\bm{G}$, \textit{i.e.},  
$g\bk=\bk+\bm{G}$ for $g\in G_{\bk}$, in this paper we consider only the wave vectors in the BZ interior, therefore $\bm{G}=\bm{0}$. 
The group $G_{\bk}$ may include rotations about $\bk$ and reflections in the planes passing through $\bk$ (and also, in the case of $\bk=\bm{0}$, inversion $I$). 
The rest of the elements of $\mathbb{G}$ form a set $Q_{\bk}=\mathbb{G}-G_{\bk}$, so that the star of $\bk$ is defined as the set of wave vectors $q\bk$, where $q\in Q_{\bk}$. 
The transformation properties of the conjugate Bloch states 
$|\bk,1\rangle$ and $|\bk,2\rangle$ under the elements of $G_{\bk}$ depend on the crystal symmetry and the direction of $\bk$, and are analyzed in the next section.

\section{The group of $\bk$ and its corepresentations}
\label{sec: G_k}

Consider a wave vector $\bk$ in the BZ interior.
The corresponding Bloch states have the form $\langle\br|\bk,s\rangle={\cal V}^{-1/2}e^{i\bk\br}\varphi_{\bk,s}(\br)$, see Eq. (\ref{Bloch-states-1-2}), where 
the lattice-periodic spinors $\varphi_{\bk,s}$ are the eigenfunctions of the reduced Hamiltonian
\begin{equation}
\label{H_k}
  \hat H_{\bk}=\frac{(\hbp+\bk)^2}{2m}+U(\br)+\frac{\hbar}{4m^2c^2}\hat{\bm{\sigma}}[\bm{\nabla}U(\br)\times(\hbp+\bk)],
\end{equation}
such that $\hat H_{\bk}\varphi_{\bk,s}=\epsilon(\bk)\varphi_{\bk,s}$. Since $\hat H_{\bk}$ is invariant under all operations from the group of $\bk$, one can classify its eigenstates according to the 
irreducible representations (irreps) of $G_{\bk}$. Using the relations ${\cal C}(\br,\hbp,\hat{\bm{\sigma}}){\cal C}^{-1}=(-\br,\hbp,-\hat{\bm{\sigma}})$ and ${\cal C}\bk{\cal C}^{-1}=\bk$, we have 
${\cal C}\hat H_{\bk}{\cal C}^{-1}=\hat H_{\bk}$, \textit{i.e.}, $\hat H_{\bk}$ is also invariant under the conjugation operation. 
Therefore, the full symmetry group of the reduced Hamiltonian at given $\bk$ is actually given by
\begin{equation}
\label{magnetic G_k}
  {\cal G}_{\bk}=G_{\bk}+{\cal C}G_{\bk},
\end{equation}
where ${\cal C}$ commutes with all elements of $G_{\bk}$. It is the additional conjugation symmetry that leads to the eigenvalues of $\hat H_{\bk}$ being at least twofold degenerate at each $\bk$. 

Since ${\cal C}$ is antiunitary, ${\cal G}_{\bk}$ is a Type II magnetic, or Shubnikov, point group, see Ref. \cite{BC-book}, and the symmetry properties of the eigenstates $\varphi_{\bk,s}$ are determined 
by the irreducible \textit{corepresentations} (coreps) of ${\cal G}_{\bk}$. The coreps of ${\cal G}_{\bk}$ can be obtained from the usual irreps of the unitary component $G_{\bk}$ 
using a standard procedure \cite{BD68,BC-book}, which is outlined in Appendix \ref{app: coreps}. 
The coreps belong to one of three cases, A, B, or C, which determine whether or not the antiunitary symmetry leads to an additional degeneracy and also the type of this degeneracy. 

An additional complication is that, since the electron wave functions are spin-1/2 spinors, any rotation by $2\pi$ changes their sign. This double-valuedness can be dealt with in the standard fashion \cite{LL-3}, 
by introducing a fictitious new symmetry element $\bar E$, which corresponds to a $2\pi$ rotation, commutes with all other elements, 
and satisfies the conditions $C_{2\bm{n}}^2=\sigma_{\bm{n}}^2=\bar E$ and $\bar E^2=E$ ($E$ is the identity element). Then, for each $G_{\bk}$ there is a corresponding double group $G'_{\bk}$, with twice as many elements. 
The physically relevant coreps of ${\cal G}_{\bk}$ are constructed from the double-valued irreps of $G_{\bk}$, which in turn are given by the single-valued irreps of $G'_{\bk}$ having the property 
$\chi(\bar E)=-\chi(E)$. Here and below $\chi(g)$ denotes the character of the group element $g$.

The Bloch states $|\bk,1\rangle$ and $|\bk,2\rangle={\cal C}|\bk,1\rangle$ form the basis of a two-dimensional (2D) corep of the magnetic group (\ref{magnetic G_k}). 
In particular, under an element $g$ of the unitary component $G_{\bk}$ we have
\begin{equation}
\label{G_k-transform-general}
  g|\bk,s\rangle=\sum_{s'}|\bk,s'\rangle{\cal D}_{s's}(g),
\end{equation}
where $\hat{\cal D}(g)$ is the corep matrix. The corep matrices for the remaining elements of ${\cal G}_{\bk}$ can be obtained by using 
\begin{equation}
\label{corep-bar E-C}
  \hat{\cal D}(\bar E)=\left(\begin{array}{cc}
                           -1 & 0 \\
                           0 & -1
                           \end{array}\right),\quad 
  \hat{\cal D}({\cal C})=\left(\begin{array}{cc}
                           0 & -1 \\
                           1 & 0
                           \end{array}\right),			   
\end{equation}
and the corep multiplication rules, see Appendix \ref{app: coreps}. The states $|\bk,1\rangle$ and $|\bk,2\rangle$ can be regarded as the pseudospin states if the corep defined by 
Eqs. (\ref{G_k-transform-general}) and (\ref{corep-bar E-C}) is equivalent to the corep spanned by the basis spinors $\xi_1$ and $\xi_2$. For the latter we have $\hat{\cal D}^{(1/2)}(g)=\hat D^{(1/2)}(g)$, 
see Eqs. (\ref{xis-R}) and (\ref{xis-sigma}), while $\hat{\cal D}^{(1/2)}(\bar E)$ and $\hat{\cal D}^{(1/2)}({\cal C})$ have the form (\ref{corep-bar E-C}).

Our procedure will be as follows. First, for each $\bk$ in the BZ interior we find the group of $\bk$ and list all its double-valued irreps. 
Then, for each double-valued irrep $\Gamma$, we apply the Dimmock-Wheeler test, Eq. (\ref{app: Dimmock-Wheeler}), to determine which corep case is realized for the corep of the magnetic group ${\cal G}_{\bk}$ 
derived from $\Gamma$. It turns out that all double-valued irreps at $\bk\neq\bm{0}$ are either one-dimensional (1D) or 2D, producing only 2D coreps, as discussed in Appendix \ref{app: 2D coreps}. 
Finally, we check if the corep $\hat{\cal D}_\Gamma$ derived from $\Gamma$ is equivalent to $\hat{\cal D}^{(1/2)}$.

\subsection{General $\bk$}
\label{sec: general k}

The simplest case is realized when $\bk$ is a general wave vector in the BZ interior, which does not have any special symmetries. In this case, the group of $\bk$ is given by $G_{\bk}=\mathbf{C}_1=\{E\}$ and the 
corresponding double group is $\mathbf{C}'_1=\{E,\bar E\}$. The only double-valued irrep of $\mathbf{C}_1$ is $\Gamma=\Gamma_2$, which is 1D. Here and below we use the notations for the double-group irreps and 
the character tables from Ref. \cite{BC-book}. The Dimmock-Wheeler formula (\ref{app: Dimmock-Wheeler}) takes the form
$$
  \sum_{g\in\mathbf{C}'_1}\chi_{\Gamma_2}(g^2)=\chi(E^2)+\chi(\bar E^2)=2=|\mathbf{C}'_1|,
$$
where $|G|$ is the order of the group $G$. The 2D corep derived from $\Gamma_2$ belongs to Case B (``doubling'' type), see Eq. (\ref{app: corep-B-1D}), and is equivalent to the spin-$1/2$ corep. 
Therefore, for a general $\bk$ the conjugate Bloch states $|\bk,1\rangle$ and $|\bk,2\rangle$ transform under ${\cal G}_{\bk}=\mathbf{C}_1+{\cal C}\mathbf{C}_1$ as the basis spinors.

\subsection{High symmetry planes}
\label{sec: planes}

Now suppose $\bk$ is in a plane of symmetry passing through the $\Gamma$ point, with the reflection in the plane denoted by $\sigma$. The group of $\bk$ is given by $G_{\bk}=\mathbf{C}_s=\{E,\sigma\}$ 
and the corresponding double group is $\mathbf{C}'_s=\{E,\sigma,\bar E,\bar\sigma\}$, where $\bar\sigma=\bar E\sigma$. There are two double-valued irreps, $\Gamma_3$ and $\Gamma_4$, both 1D, 
which are complex conjugate to each other. Therefore, one can expect that the corresponding 2D corep is Case C (``pairing'' type). Indeed, Eq. (\ref{app: Dimmock-Wheeler}) becomes
$$
  \sum_{g\in\mathbf{C}'_s}\chi_{\Gamma_3}(g^2)=\sum_{g\in\mathbf{C}'_s}\chi_{\Gamma_4}(g^2)=\chi(E^2)+\chi(\sigma^2)+\chi(\bar E^2)+\chi(\bar\sigma^2)=0.
$$
Taking $\Gamma=\Gamma_3$ and using $\chi_{\Gamma_3}(\sigma)=-i$, Eq. (\ref{app: corep-C-1D}) yields the following corep matrix: 
\begin{equation}
\label{planes-D}
  \hat{\cal D}_{\Gamma_3}(\sigma)=\left(\begin{array}{cc}
              -i & 0 \\
              0 & i
              \end{array}\right)=\hat{\cal D}^{(1/2)}(\sigma).
\end{equation}
Here we used Eq. (\ref{xis-sigma}), with the quantization axis for the basis spinors chosen along the normal to the plane.
Thus we see that if $\bk$ is in a plane of symmetry then $|\bk,1\rangle$ and $|\bk,2\rangle$ transform under the operations from ${\cal G}_{\bk}=\mathbf{C}_s+{\cal C}\mathbf{C}_s$ as the basis spinors.

\subsection{High symmetry lines}
\label{sec: lines}

We consider only the special lines passing through the $\Gamma$ point, which are denoted by $\Sigma$, $\Delta$, $\Lambda$, or $\mathrm{T}$, see Ref. \cite{BC-book} for the crystallographic nomenclature. 
It is straightforward to inspect all possible cases for the eleven centrosymmetric point groups, with the results presented in Table \ref{table: group of k - lines}. Note that, given the point group and the high symmetry line, 
changing the centering of the Bravais lattice leads to the same $G_{\bk}$ up to an isomorphism. 

\begin{table}
\caption{The groups of $\bk$ for the high symmetry lines passing through the $\Gamma$ point (rows), for all centrosymmetric point groups $\mathbb{G}$ (columns). Different crystal systems are separated by double vertical lines.}
\begin{tabular}{|c||c||c||c||c|c||c|c||c|c||c|c|}
    \hline
      \hspace*{0.6cm}   & $\ \mathbf{C}_{i}\ $& $\ \mathbf{C}_{2h}\ $ & $\ \mathbf{D}_{2h}\ $  & $\ \mathbf{C}_{4h}\ $ & $\ \mathbf{D}_{4h}\ $ & $\ \mathbf{C}_{3i}\ $ & $\ \mathbf{D}_{3d}\ $ & 
            $\ \mathbf{C}_{6h}\ $ & $\ \mathbf{D}_{6h}\ $ & $\ \mathbf{T}_{h}\ $ & $\ \mathbf{O}_{h}\ $ \\ \hline
    $\Sigma$ & - & - & $\mathbf{C}_{2v}$ & $\mathbf{C}_{s}$ & $\mathbf{C}_{2v}$ & $\mathbf{C}_{1}$ & $\mathbf{C}_{2}$ & $\mathbf{C}_{s}$ & $\mathbf{C}_{2v}$ & $\mathbf{C}_{s}$ & $\mathbf{C}_{2v}$ \\ \hline
    $\Delta$ & - & - & $\mathbf{C}_{2v}$ & $\mathbf{C}_{s}$ & $\mathbf{C}_{2v}$ & - & - & $\mathbf{C}_{6}$ & $\mathbf{C}_{6v}$ & $\mathbf{C}_{2v}$ & $\mathbf{C}_{4v}$ \\ \hline
    $\Lambda$ & - & $\mathbf{C}_{2}$ & $\mathbf{C}_{2v}$ & $\mathbf{C}_{4}$ & $\mathbf{C}_{4v}$ & $\mathbf{C}_{3}$ & $\mathbf{C}_{3v}$ & - & - & $\mathbf{C}_{3}$ & $\mathbf{C}_{3v}$ \\ \hline
    $\mathrm{T}$ & - & - & - & - & - & - & - & $\mathbf{C}_{s}$ & $\mathbf{C}_{2v}$ & - & - \\ \hline
\end{tabular}
\label{table: group of k - lines}
\end{table}

For each double-valued irrep of $G_{\bk}$, we determine the corresponding corep case and compare the corep matrices with those for the spin-1/2 basis spinors. 
The groups $G_{\bk}=\mathbf{C}_1$ and $\mathbf{C}_s$ have been considered in Secs. \ref{sec: general k} and \ref{sec: planes}, respectively, with the result that their coreps are always equivalent to the spin-$1/2$ corep.  
In the remaining cases of $G_{\bk}=\mathbf{C}_{n}$ and $\mathbf{C}_{nv}$  ($n=2,3,4$, or $6$) we choose the quantization axis for the basis spinors (the $z$ axis) to be along $\bk$.   
Then, the action of a rotation through an angle $\theta$ about $\bk$ is given by
\begin{equation}
\label{spinor-corep-R}
  \hat{\cal D}^{(1/2)}(R)=\left(\begin{array}{cc}
                           e^{-i\theta/2} & 0 \\
                           0 & e^{i\theta/2}
                           \end{array}\right).
\end{equation}
For the reflections in a ``vertical'' plane passing through $\bk$, choosing the normal to the plane along $\hat{\bm{y}}$, we obtain: 
\begin{equation}
\label{spinor-corep-sigma}
  \hat{\cal D}^{(1/2)}(\sigma_y)=\left(\begin{array}{cc}
                           0 & -1 \\
                           1 & 0
                           \end{array}\right),
\end{equation}
from Eq. (\ref{xis-sigma}).

Properties of the double-valued coreps for the high-symmetry lines are summarized in Table \ref{table: coreps-lines}. It turns out that all these coreps are 2D, which means that the Bloch bands along the special lines 
in the BZ interior are twofold degenerate, barring an accidental additional degeneracy.
We will illustrate our procedure using as an example $G_{\bk}=\mathbf{C}_{3v}$, which is realized for the $\Lambda$ lines in trigonal ($\mathbb{G}=\mathbf{D}_{3d}$) and cubic ($\mathbb{G}=\mathbf{O}_{h}$) crystals.

\begin{table}
\caption{The double-valued coreps of ${\cal G}_{\bk}$ for the high symmetry lines passing through the $\Gamma$ point. All coreps are 2D, and the third column shows the dimension of the irrep from which the corep is derived. 
The Case C coreps are derived from pairs of complex conjugate 1D irreps $(\Gamma,\Gamma^*)$. The last column shows whether the Bloch states forming the corep basis transform as the spin-1/2 states. }
\begin{tabular}{|c|c|c|c|c|}
    \hline
    $\quad G_{\bk}\quad $ & corep & \ dim $\Gamma$\ \ & \ corep case\ \ & \ pseudospin\ \ \\ \hline
    $\mathbf{C}_{1}$  & $\Gamma_2$ & 1 & B & Y  \\ \hline
    $\mathbf{C}_{s}$  & $(\Gamma_3,\Gamma_4)$  & 1 & C & Y \\ \hline
    $\mathbf{C}_{2}$  & $(\Gamma_3,\Gamma_4)$   & 1 & C & Y  \\ \hline
    $\mathbf{C}_{2v}$  & $\Gamma_5$   & 2 & A & Y \\ \hline
    $\mathbf{C}_{3}$  & $(\Gamma_4,\Gamma_5)$   & 1 & C & Y \\ 
                       & $\Gamma_6$   & 1 & B & N  \\ \hline
    $\mathbf{C}_{3v}$   & $\Gamma_4$   & 2 & A & Y  \\ 
                       & $(\Gamma_5,\Gamma_6)$  & 1 & C & N \\ \hline
    $\mathbf{C}_{4}$  & $(\Gamma_5,\Gamma_6)$  & 1 & C & Y  \\ 
                       & $(\Gamma_7,\Gamma_8)$  & 1 & C & N  \\ \hline
    $\mathbf{C}_{4v}$  & $\Gamma_6$  & 2 & A & Y  \\ 
                       & $\Gamma_7$  & 2 & A & N  \\ \hline
    $\mathbf{C}_{6}$  & $(\Gamma_7,\Gamma_8)$  & 1 & C & Y  \\ 
		      & $(\Gamma_{9},\Gamma_{10})$  & 1 & C & N  \\
                       & $(\Gamma_{11},\Gamma_{12})$  & 1 & C & N  \\ \hline
    $\mathbf{C}_{6v}$  & $\Gamma_7$   & 2 & A & Y  \\
		       & $\Gamma_8$   & 2 & A & N  \\ 
                       & $\Gamma_9$   & 2 & A & N  \\ \hline
\end{tabular}
\label{table: coreps-lines}
\end{table}

The group $\mathbf{C}_{3v}$ is generated by the rotations $C_{3z}^+$ and reflections $\sigma_y$ and has three double-valued irreps: $\Gamma_5$ and $\Gamma_6$, which are 1D and complex conjugate to each other, and also 
$\Gamma_4$, which is 2D. Taking the characters of the double group elements from Ref. \cite{BC-book} and observing that $C^{\pm,2}_{3z}=\bar C^{\mp}_{3z}=\bar EC^{\mp}_{3z}$ and $\sigma_y^2=\bar E$, 
Eq. (\ref{app: Dimmock-Wheeler}) yields 
$$
  \sum_{g\in\mathbf{C}'_{3v}}\chi(g^2)=2\left[\chi(E)+\chi(\bar C^-_{3z})+\chi(\bar C^+_{3z})+3\chi(\bar E)\right]=\left\{\begin{array}{rl}
                                      -12  & \ \mathrm{,\ for\ }\Gamma_4\\
                                      0  & \ \mathrm{,\ for\ }\Gamma_5,\Gamma_6.
                                      \end{array}\right.
$$
Therefore, $\Gamma_4$ produces a 2D corep of Case A, while $\Gamma_5$ and $\Gamma_6$ pair up to form one 2D corep of Case C. In the former case, we obtain from Eq. (\ref{app: corep-A-2D}):
$$
  \chi_{\Gamma_4}(C_{3z}^+)=1=\chi^{(1/2)}(C_{3z}^+),\quad \chi_{\Gamma_4}(\sigma_y)=0=\chi^{(1/2)}(\sigma_y),
$$
which means that the corep derived from $\Gamma_4$ is equivalent to the spin-$1/2$ corep. For the Case C corep, we choose $\Gamma=\Gamma_5$, $\Gamma^*=\Gamma_6$ in Eq. (\ref{app: corep-C-1D}) and obtain:
$$
    \hat{\cal D}_{\Gamma_5}(C^+_{3z})=\left(\begin{array}{cc}
              -1 & 0 \\
              0 & -1
              \end{array}\right),\quad
    \hat{\cal D}_{\Gamma_5}(\sigma_y)=\left(\begin{array}{cc}
              -i & 0 \\
              0 & i
              \end{array}\right).
$$
Comparing these matrices with Eqs. (\ref{spinor-corep-R}) and (\ref{spinor-corep-sigma}), we see that the $(\Gamma_5,\Gamma_6)$ corep is not equivalent to the spin-$1/2$ corep, therefore the Bloch states 
$|\bk,1\rangle$ and $|\bk,2\rangle$ do not transform under ${\cal G}_{\bk}$ as the basis spinors. It is easy to show that the basis of this corep can be chosen in the form 
$(\phi,\bar\phi)\propto(\xi_1^3+i\xi_2^3,\xi_2^3+i\xi_1^3)$. We prefer to rotate the basis by a unitary matrix $e^{-i\pi\hat\sigma_1/4}$ to obtain an equivalent corep: 
\begin{equation}
\label{C3v-Gamma_5}
    \hat{\cal D}_{\Gamma_5}(C^+_{3z})=\left(\begin{array}{cc}
              -1 & 0 \\
              0 & -1
              \end{array}\right),\quad
    \hat{\cal D}_{\Gamma_5}(\sigma_y)=\left(\begin{array}{cc}
              0 & -1 \\
              1 & 0
              \end{array}\right),
\end{equation}
for which we have $(\phi,\bar\phi)\propto(\xi_1^3,\xi_2^3)$.

\subsection{$\Gamma$ point}
\label{sec: Gamma point}

The group of $\bk$ at the $\Gamma$ point is the crystal point group $\mathbb{G}$ itself, therefore the corresponding magnetic group (\ref{magnetic G_k}) takes the form
\begin{equation}
\label{magnetic G_0}
  {\cal G}_{\bk=\bm{0}}=\mathbb{G}+{\cal C}\mathbb{G}.
\end{equation}
Since each centrosymmetric point group can be represented as a direct product of some other (noncentrosymmetric) point group $\tilde{\mathbb{G}}$ and $\mathbf{C}_i=\{E,I\}$, the last expression can also be written as 
${\cal G}_{\bk=\bm{0}}=\mathbb{G}+K\mathbb{G}$.

The Bloch states at the $\Gamma$ point transform according to the double-valued coreps of ${\cal G}_{\bk=\bm{0}}$, which are obtained using the procedure described in Appendix \ref{app: coreps}. 
The results are shown in Table \ref{table: coreps-point}. 
Most of the double-valued coreps are 2D, leading to the electron bands being twofold degenerate at the $\Gamma$ point. There are just two exceptions, which are four-dimensional (4D), both in the cubic 
system: (i) the Case C corep derived from the pair of 2D irreps $(\Gamma_6,\Gamma_7)$ of $\tilde{\mathbb{G}}=\mathbf{T}$ and (ii) the Case A corep derived from the 4D irrep $\Gamma_8$ of $\tilde{\mathbb{G}}=\mathbf{O}$.

The irreps of $\mathbb{G}$ and therefore the coreps of ${\cal G}_{\bk=\bm{0}}$ are either even ($\Gamma^+$) or odd ($\Gamma^-$) under inversion. In Table \ref{table: coreps-point}, 
we use $\mathbb{G}=\tilde{\mathbb{G}}\times\mathbf{C}_i$ and list the irreps $\Gamma$ of the point group 
$\tilde{\mathbb{G}}$, with the understanding that each element of $\mathbb{G}$ has the form $g=\tilde g$ or $g=I\tilde g$, where $\tilde g\in\tilde{\mathbb{G}}$. Therefore, the corep matrices are given by 
\begin{equation}
\label{Gamma-pm}
  \hat{\cal D}_{\Gamma^\pm}(\tilde g)=\hat{\cal D}_{\Gamma}(\tilde g),\quad \hat{\cal D}_{\Gamma^\pm}(I\tilde g)=\pm\hat{\cal D}_{\Gamma}(\tilde g).
\end{equation}
The inversion-odd 2D coreps and all 4D coreps cannot be equivalent to the spin-1/2 representation. 

As an example, let us consider the point group $\mathbf{T}_{h}=\mathbf{T}\times\mathbf{C}_i$. The group $\tilde{\mathbb{G}}=\mathbf{T}$ has three double-valued irreps, all 2D: 
$\Gamma_5$, $\Gamma_6$, and $\Gamma_7$, the last two being complex 
conjugate to each other. Taking the characters of the double group elements from Ref. \cite{BC-book} and using $C^{\pm,2}_3=\bar C^{\pm,2}_3=\bar C^{\mp}_3$ and $C_2^2=\bar E$, we obtain from
Eq. (\ref{app: Dimmock-Wheeler}): 
$$
  \sum_{g\in\mathbf{T}'}\chi(g^2)=2\left[\chi(E)+4\chi(\bar C^-_3)+4\chi(\bar C^+_3)+3\chi(\bar E)\right]=\left\{\begin{array}{rl}
                                      -24  & \ \mathrm{,\ for\ }\Gamma_5\\
                                      0  & \ \mathrm{,\ for\ }\Gamma_6,\Gamma_7.
                                      \end{array}\right.
$$
The inversion-even ($\Gamma^+_5$) and inversion-odd ($\Gamma^-_5$) coreps of ${\cal G}_{\bk=\bm{0}}$ derived from $\Gamma_5$ belong to Case A and are 2D, see Eq. (\ref{app: corep-A-2D}). 
Comparing with the characters of rotations in the spin-$1/2$ representation, see Eq. (\ref{spinor-corep-R}), we have $\chi_{\Gamma^\pm_5}(C_3)=1=\chi^{(1/2)}(C_3)$, 
$\chi_{\Gamma^\pm_5}(C_2)=0=\chi^{(1/2)}(C_2)$, and $\chi_{\Gamma^\pm_5}(I)=\pm 2=\pm\chi^{(1/2)}(I)$. 
Therefore, $\Gamma^+_5$ is equivalent to the spin-1/2 representation. In contrast, the irreps $\Gamma_6$ and $\Gamma_7$ pair up to form a single 4D corep of Case C, whose basis does not transform as spin-$1/2$ spinors.

\begin{table}
\caption{The double-valued coreps of ${\cal G}_{\bk=\bm{0}}=\mathbb{G}+{\cal C}\mathbb{G}$. Each centrosymmetric point group has the form $\mathbb{G}=\tilde{\mathbb{G}}\times\mathbf{C}_i$. 
The double-valued irreps of $\tilde{\mathbb{G}}$ from which the coreps of ${\cal G}_{\bk=\bm{0}}$ are derived and the irrep dimensions are listed in the third and fourth columns, respectively.
The Case C coreps are derived from pairs of complex conjugate 1D irreps $(\Gamma,\Gamma^*)$. 
All coreps are 2D, except the $(\Gamma_6^\pm,\Gamma_7^\pm)$ coreps for $\mathbb{G}=\mathbf{T}_{h}$ and the $\Gamma_8^\pm$ coreps for $\mathbb{G}=\mathbf{O}_{h}$, which are 4D.
The last column shows whether the basis of an inversion-even corep ($\Gamma^+$) transforms as spin-1/2 states (the $\Gamma^+_8$ corep of $\mathbf{O}_{h}$ is equivalent to the spin-$3/2$ representation \cite{Lax-book}).}
\begin{tabular}{|c|c|c|c|c|c|}
    \hline
    $\quad \mathbb{G}\quad $ & $\quad \tilde{\mathbb{G}}\quad $ & corep & \ dim $\Gamma $\ \ & \ corep case\ \ & \ pseudospin\ \ \\ \hline
    $\mathbf{C}_{i}$  & $\mathbf{C}_{1}$ & $\Gamma_2$ & 1 & B & Y  \\ \hline
    $\mathbf{C}_{2h}$  & $\mathbf{C}_{2}$ & $(\Gamma_3,\Gamma_4)$ & 1 & C & Y \\ \hline
    $\mathbf{D}_{2h}$  & $\mathbf{D}_{2}$ & $\Gamma_5$  & 2 & A & Y  \\ \hline
    $\mathbf{C}_{4h}$  & $\mathbf{C}_{4}$ & $(\Gamma_5,\Gamma_6)$  & 1 & C & Y \\ 
                       & & $(\Gamma_7,\Gamma_8)$  & 1 & C & N  \\ \hline
    $\mathbf{D}_{4h}$  & $\mathbf{D}_{4}$ & $\Gamma_6$  & 2 & A & Y \\ 
		       & & $\Gamma_7$  & 2 & A & N \\ \hline
    $\mathbf{C}_{3i}$  & $\mathbf{C}_{3}$ & $(\Gamma_4,\Gamma_5)$  & 1 & C & Y  \\ 
                       & & $\Gamma_6$ & 1 & B & N  \\ \hline		       
    $\mathbf{D}_{3d}$  & $\mathbf{D}_{3}$ & $\Gamma_4$  & 2 & A & Y  \\ 
		      & & $(\Gamma_5,\Gamma_6)$ & 1 & C & N  \\ 
                       \hline
    $\mathbf{C}_{6h}$  & $\mathbf{C}_{6}$ & $(\Gamma_7,\Gamma_8)$  & 1 & C & Y  \\ 
		      & & $(\Gamma_9,\Gamma_{10})$  & 1& C & N  \\
                       & & $(\Gamma_{11},\Gamma_{12})$  & 1 & C & N  \\ \hline			
    $\mathbf{D}_{6h}$  & $\mathbf{D}_{6}$ & $\Gamma_7$ & 2 & A & Y  \\ 
                       & & $\Gamma_8$ & 2 & A & N  \\ 
                       & & $\Gamma_9$ & 2 & A & N  \\ \hline
    $\mathbf{T}_{h}$  & $\mathbf{T}$ & $\Gamma_5$  & 2 & A & Y  \\
		       & & $(\Gamma_6,\Gamma_7)$  & 2 & C & N  \\ \hline                       
    $\mathbf{O}_{h}$  & $\mathbf{O}$ & $\Gamma_6$  & 2 & A & Y  \\
		       & & $\Gamma_7$  & 2 & A & N  \\ 
                       & & $\Gamma_8$  & 4 & A & N  \\ \hline		
\end{tabular}
\label{table: coreps-point}
\end{table}

\section{Bloch basis for the whole BZ}
\label{sec: whole-BZ}

It follows from the analysis in the previous section that the pseudospin representation fails for some coreps at the $\Gamma$ point and along the high-symmetry lines, \textit{i.e.},  the Bloch states 
$|\bk,1\rangle$ and $|\bk,2\rangle={\cal C}|\bk,1\rangle$ transform under ${\cal G}_{\bk}$ according to Eq. (\ref{G_k-transform-general}), but $\hat{\cal D}(g)$ is not equivalent to $\hat{\cal D}^{(1/2)}(g)$. 
These exceptional coreps are indicated in the last columns of Tables \ref{table: coreps-lines} and \ref{table: coreps-point}. 
If, in a given band, the Bloch states for all $\bk$ in the BZ interior correspond to the coreps of ${\cal G}_{\bk}$ equivalent to the spin-1/2 corep,
then the band is called ``pseudospin band''. If the Bloch states at the $\Gamma$ point correspond to a corep which is not equivalent to the spin-$1/2$ corep, then the pseudospin 
representation also fails along the high-symmetry lines, due to the compatibility relations, see below. In this case, the band is called ``non-pseudospin band''. 

Let us first look at the case of a pseudospin band. For any $\bk$ we have  
\begin{equation}
\label{G_k-transform-pseudospin}
  g|\bk,s\rangle=\sum_{s'}|\bk,s'\rangle D^{(1/2)}_{s's}(g),
\end{equation}
under $g\in G_{\bk}$. In order to define the Bloch bases across the whole BZ in such a way that $|\bk,1\rangle$ and $|\bk,2\rangle$ transform like the basis spinors under all operations from $\mathbb{G}$, 
we start with some $\bk$ in the fundamental domain of the BZ and apply an element $q\in Q_{\bk}$ to transform $\bk$ into $q\bk$ -- a ray of the star of $\bk$. Since the state $q|\bk,s\rangle$ belongs to 
the wave vector $q\bk$, it can be represented as
\begin{equation}
\label{star-k-transform}
  q|\bk,s\rangle=\sum_{s'}|q\bk,s'\rangle U_{\bk,s's}(q),
\end{equation}
where the expansion coefficients form a unitary matrix. Following Ref. \cite{UR85}, we choose this matrix in the form $\hat U_{\bk}(q)=\hat D^{(1/2)}(q)$ and use the expressions
\begin{equation}
\label{Ueda-Rice-h}
    q|\bk,s\rangle=\sum_{s'}|q\bk,s'\rangle D_{s's}^{(1/2)}(q)
\end{equation}
to \textit{define} the pseudospin bases $(|q\bk,1\rangle,|q\bk,2\rangle)$ for the whole star of $\bk$. Combining Eqs. (\ref{G_k-transform-pseudospin}) and (\ref{Ueda-Rice-h}), we obtain
the Ueda-Rice formula:
\begin{equation}
\label{Ueda-Rice}
    g|\bk,s\rangle=\sum_{s'}|g\bk,s'\rangle D_{s's}^{(1/2)}(g),\quad g\in\mathbb{G}.
\end{equation}
In particular, $I|\bk,s\rangle=|-\bk,s\rangle$.

It is easy to see that the expression (\ref{Ueda-Rice}) cannot work in a non-pseudospin band, because it is not compatible with the transformation properties of the Bloch states at the special locations in the BZ. 
Indeed, assuming the Bloch basis continuity, Eq. (\ref{Ueda-Rice}) yields $g|\bm{0},s\rangle=\sum_{s'}|\bm{0},s'\rangle D_{s's}^{(1/2)}(g)$, which is not consistent with the fact that the corep at 
the $\Gamma$ point is not necessarily equivalent to the spin-$1/2$ corep. This continuity argument suggests a natural generalization of the Ueda-Rice prescription. Suppose the Bloch states at the 
$\Gamma$ point transform according to a 2D corep $\Gamma$ of ${\cal G}_{\bk=\bm{0}}$, then the Bloch basis in the whole BZ can be defined by the following expression:
\begin{equation}
\label{general-prescription}
    g|\bk,s\rangle=\sum_{s'}|g\bk,s'\rangle {\cal D}_{\Gamma,s's}(g),\quad g\in\mathbb{G}.
\end{equation}
In particular, 
\begin{equation}
\label{general-prescription-I}
    I|\bk,s\rangle=p_\Gamma|-\bk,s\rangle,
\end{equation}
where $p_\Gamma=\pm$ denotes the corep parity. Under TR operation $K={\cal C}I$, we have 
\begin{equation}
\label{general-prescription-K}
    K|\bk,1\rangle=p_\Gamma|-\bk,2\rangle,\quad K|\bk,2\rangle=-p_\Gamma|-\bk,1\rangle.
\end{equation}
The matrices of all 2D non-pseudospin coreps are given in Table \ref{table: Gamma-point-corep-matrices}. 
Note that the prescription (\ref{general-prescription}) is not applicable for the bands which are fourfold degenerate at the $\Gamma$ point, 
namely the $(\Gamma_6^\pm,\Gamma_7^\pm)$ bands for $\mathbb{G}=\mathbf{T}_{h}$ and the $\Gamma_8^\pm$ ($j=3/2$) bands for $\mathbb{G}=\mathbf{O}_{h}$. These cases require a different treatment, see, \textit{e.g.}, 
Ref. \cite{Lutt56}, and will not be considered here.

Since each centrosymmetric point group $\mathbb{G}$ is generated by the generators of $\tilde{\mathbb{G}}$ 
and also by inversion $I$, one can use Eq. (\ref{Gamma-pm}) and Table \ref{table: Gamma-point-corep-matrices} to obtain $\hat{\cal D}_\Gamma(g)$ for all $g\in \mathbb{G}$. 
For example, the point group $\mathbb{G}=\mathbf{D}_{4h}$ has two double-valued non-pseudospin coreps of opposite parity, $\Gamma_7^+$ and $\Gamma_7^-$, derived from the irrep $\Gamma_7$ of $\tilde{\mathbb{G}}=\mathbf{D}_4$. 
The latter group has two generators, $\tilde g_1=C_{4z}^+$ and $\tilde g_2=C_{2y}$, and we obtain:
\begin{eqnarray*}
  && \hat{\cal D}_{\Gamma_7^\pm}(C_{4z}^+)=\left(\begin{array}{cc}
                                                               -e^{-i\pi/4} & 0 \\
                                                               0 & -e^{i\pi/4}
                                                               \end{array}\right)=-\hat D^{(1/2)}(C^+_{4z}),\\
  && \hat{\cal D}_{\Gamma_7^\pm}(C_{2y})=\left(\begin{array}{cc}
                                                               0 & -1 \\
                                                               1 & 0
                                                               \end{array}\right)=\hat D^{(1/2)}(C_{2y}),                                                         
\end{eqnarray*}
and $\hat{\cal D}_{\Gamma_7^\pm}(I)=\pm\hat\sigma_0$.

\begin{table}
\caption{The corep matrices for the non-pseudospin 2D coreps of ${\cal G}_{\bk=\bm{0}}$, with $\tilde g$ denoting the generators of $\tilde{\mathbb{G}}$.
The Case C coreps are derived from pairs of complex conjugate 1D irreps $(\Gamma,\Gamma^*)$. Only the inversion-even bases ($\phi,\bar\phi={\cal C}\phi$) are shown, 
and $f(\br)$ in the last row is the basis function of the $\Gamma^+_2$ irrep of $\mathbf{O}_h$ (changing sign under a $C_{4z}^+$ rotation), \textit{e.g.}, $f=x^4(y^2-z^2)+y^4(z^2-x^2)+z^4(x^2-y^2)$ \cite{Lax-book}.}
\begin{tabular}{|c|c|c|c|}
    \hline
    $\quad \tilde{\mathbb{G}}\quad $ &  corep & $\hat{\cal D}_\Gamma(\tilde g)$ & $(\phi,\bar\phi)$  \\ \hline
    $\mathbf{C}_{4}$  & $(\Gamma_7,\Gamma_8)$   &  $\hat{\cal D}(C^+_{4z})=-\hat D^{(1/2)}(C^+_{4z})$ & ($\xi_2^3,-\xi_1^3$) \\ \hline
    $\mathbf{D}_{4}$ & $\Gamma_7$  & \ $\hat{\cal D}(C^+_{4z})=-\hat D^{(1/2)}(C^+_{4z})$,\quad $\hat{\cal D}(C_{2y})=\hat D^{(1/2)}(C_{2y})$\ \  & ($\xi_2^3,-\xi_1^3$) \\ \hline
    $\mathbf{C}_{3}$  & $\Gamma_6$ &  $\hat{\cal D}(C^+_{3z})=-\hat\sigma_0$ & ($\xi_1^3,\xi_2^3$) \\ \hline
    $\mathbf{D}_{3}$  & $(\Gamma_5,\Gamma_6)$  & $\hat{\cal D}(C^+_{3z})=-\hat\sigma_0$,\quad $\hat{\cal D}(C_{2y})=\hat D^{(1/2)}(C_{2y})$ & ($\xi_1^3,\xi_2^3$) \\ \hline								
    $\mathbf{C}_{6}$  & $(\Gamma_{9},\Gamma_{10})$  & $\hat{\cal D}(C^+_{6z})=-\hat D^{(1/2)}(C^+_{6z})$ & ($\xi_2^5,-\xi_1^5$) \\
                      & $(\Gamma_{11},\Gamma_{12})$  & $\hat{\cal D}(C^+_{6z})=i\hat\sigma_3$ & ($\xi_2^3,-\xi_1^3$) \\ \hline
    $\mathbf{D}_{6}$  & $\Gamma_8$  & $\hat{\cal D}(C^+_{6z})=-\hat D^{(1/2)}(C^+_{6z})$,\quad $\hat{\cal D}(C_{2y})=\hat D^{(1/2)}(C_{2y})$ & ($\xi_2^5,-\xi_1^5$) \\ 
                       & $\Gamma_9$ & $\hat{\cal D}(C^+_{6z})=i\hat\sigma_3$,\quad $\hat{\cal D}(C_{2y})=\hat D^{(1/2)}(C_{2y})$ & ($\xi_2^3,-\xi_1^3$)  \\ \hline
    $\mathbf{O}$  & $\Gamma_7$ &  $\hat{\cal D}(C^+_{4z})=-\hat D^{(1/2)}(C^+_{4z})$,\quad $\hat{\cal D}(C_{2y})=\hat D^{(1/2)}(C_{2y})$ & \ \ $f(\br)(\xi_1,\xi_2)$\ \ \ \\ 
                                     & &  $\hat{\cal D}(C^+_{3xyz})=\hat D^{(1/2)}(C^+_{3xyz})$ & \\ \hline                                                               
\end{tabular}
\label{table: Gamma-point-corep-matrices}
\end{table}

It is straightforward to check that the prescription (\ref{general-prescription}) satisfies all necessary consistency requirements. In particular, it preserves the conjugation relations between the Bloch states. 
Since $\langle\bk,s|{\cal C}|\bk,s'\rangle=-i\sigma_{2,ss'}$ and ${\cal C}$ is an antilinear operation commuting with all point group operations, we obtain:
$$
  \langle g\bk,s|{\cal C}|g\bk,s'\rangle=\sum_{s_1s_2}{\cal D}_{\Gamma,ss_1}(g){\cal D}^\top_{\Gamma,s_2s'}(g)\langle\bk,s_1|{\cal C}|\bk,s_2\rangle
    =-i[\hat{\cal D}_{\Gamma}(g)\hat\sigma_2\hat{\cal D}^\top_{\Gamma}(g)]_{ss'}=-i\sigma_{2,ss'}.
$$
Here we used the fact that the corep matrices $\hat{\cal D}_{\Gamma}$ are special unitary matrices. 
Also, the transformation properties of the global Bloch bases constructed according to Eq. (\ref{general-prescription}) are compatible with the coreps of 
${\cal G}_{\bk}$ for $\bk$ along the high symmetry lines, see Sec. \ref{sec: lines}. For those special wave vectors, Eq. (\ref{general-prescription}) takes the form
$$
\label{non-pseudospin-basis-special-k}
    g|\bk,s\rangle=\sum_{s'}|\bk,s'\rangle {\cal D}_{\Gamma,s's}(g),\quad g\in G_{\bk}.
$$
Since ${\cal G}_{\bk}$ is a subgroup of ${\cal G}_{\bk=\bm{0}}$, the matrices ${\cal D}_\Gamma$ here provide a subduced corepresentation of ${\cal G}_{\bk}$, which should be compared with the irreducible coreps listed in 
Table \ref{table: coreps-lines}. Since the pseudospin coreps are evidently compatible at all $\bk$, it is sufficient to examine only the exceptional coreps. 
The resulting compatibility relations are given in Table \ref{table: compatibility-relations}.

\begin{table}
\caption{The compatibility relations between the 2D non-pseudospin coreps of ${\cal G}_{\bk=\bm{0}}$ and the coreps of ${\cal G}_{\bk}$ for the high symmetry lines.}
\begin{tabular}{|c|c|c|c|c|c|}
    \hline
    $\quad \mathbb{G}\quad $  & $\Gamma$ point & $\Sigma$ line & $\Delta$ line & $\Lambda$ line & $\mathrm{T}$ line \\ \hline
    $\mathbf{C}_{4h}$  & $(\Gamma_7,\Gamma_8)$  & \ $(\Gamma_3,\Gamma_4)$\ \  & $(\Gamma_3,\Gamma_4)$ & \ $(\Gamma_7,\Gamma_8)$\ \  & - \\ \hline
    $\mathbf{D}_{4h}$  & $\Gamma_7$  & $\Gamma_5$ & $\Gamma_5$ & $\Gamma_7$  & - \\ \hline
    $\mathbf{C}_{3i}$  & $\Gamma_6$  & $\Gamma_2$ & - & $\Gamma_6$ & - \\ \hline
    $\mathbf{D}_{3d}$  & $(\Gamma_5,\Gamma_6)$ & $(\Gamma_3,\Gamma_4)$ & - & $(\Gamma_5,\Gamma_6)$ & - \\ \hline
    $\mathbf{C}_{6h}$  & $(\Gamma_9,\Gamma_{10})$  & $(\Gamma_3,\Gamma_4)$ & $(\Gamma_9,\Gamma_{10})$ & - & $(\Gamma_3,\Gamma_4)$  \\
                       & \ $(\Gamma_{11},\Gamma_{12})$\ \   & $(\Gamma_3,\Gamma_4)$ & \ $(\Gamma_{11},\Gamma_{12})$\ \  & - & \ $(\Gamma_3,\Gamma_4)$\ \  \\ \hline			
    $\mathbf{D}_{6h}$  & $\Gamma_8$ & $\Gamma_5$ & $\Gamma_8$ & - & $\Gamma_5$ \\ 
                       & $\Gamma_9$ & $\Gamma_5$ & $\Gamma_9$ & - & $\Gamma_5$ \\ \hline                   
    $\mathbf{O}_{h}$  & $\Gamma_7$  & $\Gamma_5$ & $\Gamma_7$ & $\Gamma_4$ & - \\ \hline		
\end{tabular}
\label{table: compatibility-relations}
\end{table}

\section{Antisymmetric spin-orbit coupling}
\label{sec: ASOC}

As an application of the theory developed above, in this section we derive the effective model of the SOC in a \textit{noncentrosymmetric} crystal. The lattice potential in Eq. (\ref{general H}) can be represented as 
$U(\br)=U_s(\br)+U_a(\br)$, where
$$
    U_s(\br)=\frac{U(\br)+U(-\br)}{2}, \quad U_a(\br)=\frac{U(\br)-U(-\br)}{2}.
$$
The Hamiltonian then takes the form $\hat H=\hat H_s+\hat H_a$, where
\begin{eqnarray}
\label{Hs}
        &&\hat H_s=\frac{\hbp^2}{2m}+U_s(\br)
        +\frac{\hbar}{4m^2c^2}\hat{\bm{\sigma}}[\bm{\nabla}U_s(\br)\times\hbp],\\
\label{Ha}
    &&\hat H_a=U_a(\br)+\frac{\hbar}{4m^2c^2}\hat{\bm{\sigma}}[\bm{\nabla}U_a(\br)\times\hbp]
\end{eqnarray}
are the inversion-symmetric and inversion-antisymmetric parts, respectively.
Proceeding as in Sec. \ref{sec: band structure}, we diagonalize Eq. (\ref{Hs}) and obtain twofold degenerate bands labelled by $s=1,2$: $\hat H_s|\bk,n,s\rangle=\epsilon_n(\bk)|\bk,n,s\rangle$. 

Both the potential $U(\br)$ and its antisymmetric part $U_a(\br)$ are invariant under the same set of proper and improper rotations forming the point group $\mathbb{G}$, 
which is one of the twenty one noncentrosymmetric point groups. However, the symmetric part $U_s(\br)$ and, therefore, $\hat H_s$ are invariant under a larger group
\begin{equation}
\label{G_s}
  \mathbb{G}_s=\mathbb{G}\times\{E,I\},
\end{equation}
which is one of the eleven centrosymmetric point groups. At the $\Gamma$ point, the conjugate Bloch states $|\bm{0},n,1\rangle$ and $|\bm{0},n,2\rangle$, see Eq. (\ref{Bloch pseudospinors}), 
form the basis of a 2D corep $\Gamma$ of the magnetic group ${\cal G}_{\bk=\bm{0}}=\mathbb{G}_s+{\cal C}\mathbb{G}_s$. 
Then the relative ``orientations'' of the Bloch bases at different $\bk$ points in each band are determined by the prescription (\ref{general-prescription}), with $g\in\mathbb{G}_s$.

Let us now calculate the matrix elements of the inversion-antisymmetric part (\ref{Ha}) in the basis of the eigenstates of $\hat H_s$. It is easy to show that $\hat H_a$ is diagonal in $\bk$ and one can write
\begin{equation}
\label{H_a-matrix-elements}
    \langle\bk,n,s|\hat H_a|\bk,n',s'\rangle=iA_{nn'}(\bk)\delta_{ss'}+\bB_{nn'}(\bk)\bm{\sigma}_{ss'}.
\end{equation}
Therefore, the general second-quantized Hamiltonian of the band electrons has the following form:
\begin{equation}
\label{H-second-quant}
  \hat H=\sum_{\bk,nn',ss'}[\epsilon_n(\bk)\delta_{nn'}\delta_{ss'}+iA_{nn'}(\bk)\delta_{ss'}+\bm{B}_{nn'}(\bk)\bm{\sigma}_{ss'}]\hat a^\dagger_{\bk ns}\hat a_{\bk n's'},
\end{equation}
where the last two terms contain all effects of the inversion symmetry breaking.  

The matrices $A$ and $\bB$ must satisfy a number of symmetry-imposed constraints. From the Hermiticity of $\hat H_a$ we obtain:
\begin{equation}
\label{AB-constraint-Hermiticity}
  A_{nn'}(\bk)=-A^*_{n'n}(\bk),\quad \bB_{nn'}(\bk)=\bB^*_{n'n}(\bk).
\end{equation}
Since $I\hat H_aI^{-1}=-\hat H_a$, we have 
$$
  \langle\bk,n,s|\hat H_a|\bk,n',s'\rangle=-\langle\bk,n,s|I^\dagger\hat H_aI|\bk,n',s'\rangle=-p_np_{n'}\langle-\bk,n,s|\hat H_a|-\bk,n',s'\rangle,
$$
where $p_n$ is the parity of the $\Gamma$-point corep in the $n$th band, see Eq. (\ref{general-prescription-I}). Therefore,
\begin{equation}
\label{AB-constraint-I}
  A_{nn'}(\bk)=-p_np_{n'}A_{nn'}(-\bk),\quad \bB_{nn'}(\bk)=-p_np_{n'}\bB_{nn'}(-\bk).
\end{equation}
It follows from the TR invariance, $K\hat H_aK^{-1}=\hat H_a$, that
\begin{eqnarray*}
  && \langle\bk,n,s|\hat H_a|\bk,n',s'\rangle = \langle\bk,n,s|K^\dagger\hat H_aK|\bk,n',s'\rangle \\
      && = -p_np_{n'}\sum_{s_1s_2}\sigma_{2,s's_1}\sigma_{2,ss_2}\langle-\bk,n',s_1|\hat H_a|-\bk,n,s_2\rangle,
\end{eqnarray*}
where we used Eq. (\ref{general-prescription-K}) and the property $\langle i|K^\dagger|j\rangle=\langle j|K|i\rangle$, which reflects the antiunitarity of $K$. Therefore,
\begin{equation}
\label{AB-constraint-K}
  A_{nn'}(\bk)=p_np_{n'}A_{n'n}(-\bk),\quad \bB_{nn'}(\bk)=-p_np_{n'}\bB_{n'n}(-\bk).
\end{equation}
From Eqs. (\ref{AB-constraint-Hermiticity}), (\ref{AB-constraint-I}), and (\ref{AB-constraint-K}) we obtain that $A_{nn'}$ and $\bB_{nn'}$ are real and satisfy $A_{nn'}(\bk)=-A_{n'n}(\bk)$ 
and $\bB_{nn'}(\bk)=\bB_{n'n}(\bk)$. Furthermore, $A$ and $\bB$ are odd (even) in $\bk$ if the bands $n$ and $n'$ have the same (opposite) parity.
The symmetry under rotations and reflections from the crystal point group imposes additional constraints, which are examined below in the cases of one and two twofold degenerate bands.

\subsection{One-band Rashba model}
\label{sec: one-band Rashba}

Keeping just one band and observing that $A_{nn}(\bk)=0$, Eq. (\ref{H_a-matrix-elements}) takes the form $\langle\bk,n,s|\hat H_a|\bk,n,s'\rangle=\bgam_n(\bk)\bm{\sigma}_{ss'}$, 
where $\bgam_n(\bk)=\bB_{nn}(\bk)$. Dropping the band index $n$, we arrive at
\begin{equation}
\label{H-a-one-band}
    \langle\bk,s|\hat H_a|\bk,s'\rangle=\bgam(\bk)\bm{\sigma}_{ss'},
\end{equation}
where $\bgam(\bk)=-\bgam(-\bk)$ is a real pseudovector. Thus we obtain the following effective Hamiltonian:
\begin{equation}
\label{Rashba-H}
  \hat H=\sum_{\bk,ss'}\left[\epsilon(\bk)\delta_{ss'}+\bgam(\bk)\bm{\sigma}_{ss'}\right]\hat a^\dagger_{\bk s}\hat a_{\bk s'},
\end{equation}
which is called the generalized Rashba model. In the original Rashba model, see Refs. \cite{Rashba60,Rashba-model-review}, the particular case with $\bgam(\bk)=\gamma_0(k_y\hat{\bm x}-k_x\hat{\bm y})$ 
was used to describe the antisymmetric SOC in quasi-2D semiconductors. 

For any element $g$ of the noncentrosymmetric point group $\mathbb{G}$, we have $g\hat H_ag^{-1}=\hat H_a$. On the other hand, since $g$ is also an element of $\mathbb{G}_s$, the eigenstates $|\bk,1\rangle$ 
and $|\bk,2\rangle$ of $\hat H_s$ transform under $g$ according to Eq. (\ref{Ueda-Rice}) in a pseudospin band, or according to Eq. (\ref{general-prescription}) in a general band. 
Then, it follows from Eq. (\ref{H-a-one-band}) that
\begin{eqnarray}
\label{H_a-matrix-transform}
  \bgam(\bk)\bm{\sigma}_{ss'}=\langle\bk,s|g^\dagger \hat H_a g|\bk,s'\rangle=\sum_{s_1s_2}{\cal D}_{\Gamma,s_1s}^*(g){\cal D}_{\Gamma,s_2s'}(g)\langle g\bk,s_1|\hat H_a|g\bk,s_2\rangle\nonumber\\
  =\bgam(g\bk)\left[\hat{\cal D}_\Gamma^\dagger(g)\hat{\bm{\sigma}}\hat{\cal D}_\Gamma(g)\right]_{ss'},
\end{eqnarray}
where $\hat{\cal D}_\Gamma(g)$ is the $\Gamma$-point corep of $\mathbb{G}_s$ subduced to $\mathbb{G}$. Using the fact that
\begin{equation}
\label{sigma-rotation-non-pseudospin}
  \hat{\cal D}_\Gamma^\dagger(g)\hat{\sigma}_i\hat{\cal D}_\Gamma(g)=\sum_{j=1}^3{\cal R}_{ij}(g)\hat{\sigma}_j,
\end{equation}
where $\hat{\cal R}$ is a $3\times 3$ orthogonal matrix, we obtain from Eq. (\ref{H_a-matrix-transform}) the following point-group invariance condition for the antisymmetric SOC:
\begin{equation}
\label{gamma-transform-gen}
  \gamma_i(\bk)=\sum_{j=1}^3{\cal R}_{ij}(g)\gamma_j(g^{-1}\bk),\quad g\in\mathbb{G}.
\end{equation}
Note that this condition does not depend on the parity of the $\Gamma$ corep, since $\hat{\cal D}_{\Gamma^+}$ and $\hat{\cal D}_{\Gamma^-}$ produce the the same ${\cal R}$ matrix.

Diagonalizing the Hamiltonian (\ref{Rashba-H}), one obtains two bands $\xi_\lambda(\bk)=\epsilon(\bk)+\lambda|\bgam(\bk)|$, where $\lambda=\pm$ is called ``helicity''. The bands are split almost everywhere, except
at the $\Gamma$ point and possibly some other high-symmetry locations in the BZ, where $\bgam(\bk)=\bm{0}$. It follows from Eq. (\ref{gamma-transform-gen}) and the property $\bgam(\bk)=-\bgam(-\bk)$ that, 
regardless of the form of $\bgam(\bk)$, the helicity band dispersions are invariant under all operations from the group $\mathbb{G}_s$.

\subsubsection{Pseudospin band}
\label{sec: Rashba-pseudospin}

In a pseudospin band, we use $\hat{\cal D}_\Gamma(g)=\hat D^{(1/2)}(R)$ in Eq. (\ref{sigma-rotation-non-pseudospin}). From the well-known expression
\begin{equation}
\label{D-to-R}
  \hat{D}^{(1/2),\dagger}(R)\hat{\bm{\sigma}}\hat{D}^{(1/2)}(R)=R\hat{\bm{\sigma}},
\end{equation}
which holds for both proper ($g=R$) and improper ($g=IR$) rotations, we obtain $\hat{\cal R}(g)=\hat R$, where $\hat R$ is the $3\times 3$ rotation matrix.
Therefore, the constraint (\ref{gamma-transform-gen}) takes the following form: 
\begin{equation}
\label{gamma-transform-pseudospin}
  \bgam(\bk)=\left\{\begin{array}{ll}
                    R\bgam(R^{-1}\bk), &\quad g=R,\\
                    -R\bgam(R^{-1}\bk), &\quad g=IR.
                    \end{array}\right.
\end{equation}
Representative expressions for the antisymmetric SOC in the vicinity of the $\Gamma$ point satisfying these conditions are given in Table \ref{table: gammas-pseudospin}, 
see also Ref. \cite{Sam09}. It should be noted that the conditions (\ref{gamma-transform-pseudospin}) are applicable for all electron bands in triclinic 
($\mathbb{G}=\mathbf{C}_1$), monoclinic ($\mathbb{G}=\mathbf{C}_2,\mathbf{C}_s$), and orthorhombic ($\mathbb{G}=\mathbf{D}_2,\mathbf{C}_{2v}$) crystals.

\begin{table}
\caption{The antisymmetric SOC near the $\Gamma$ point in the pseudospin bands; 
$a_i$ and $a$ are real constants, $b_i$ and $b$ are complex constants, and $k_\pm=k_x\pm ik_y$. For each noncentrosymmetric point group $\mathbb{G}$, the corresponding group $\mathbb{G}_s$ and its pseudospin coreps 
at the $\Gamma$ point are listed in the second and third columns, respectively.}
\begin{tabular}{|c|c|c|c|}
    \hline
    $\quad \mathbb{G}\quad $ & $\mathbb{G}_s$ & $\Gamma$ & $\bgam(\bk)$ \\ \hline
    $\mathbf{C}_1$  & $\mathbf{C}_i$ & $\Gamma_2$   & $(a_1k_x+a_2k_y+a_3k_z)\hat{\bm x}+(a_4k_x+a_5k_y+a_6k_z)\hat{\bm y}+(a_7k_x+a_8k_y+a_9k_z)\hat{\bm z}$ \\ \hline
    $\mathbf{C}_2$   & \ $\mathbf{C}_{2h}$\ \  & $(\Gamma_3,\Gamma_4)$   & $(a_1k_x+a_2k_y)\hat{\bm x}+(a_3k_x+a_4k_y)\hat{\bm y}+a_5k_z\hat{\bm z}$ \\ \hline
    $\mathbf{C}_s$   & $\mathbf{C}_{2h}$ & $(\Gamma_3,\Gamma_4)$   & $a_1k_z\hat{\bm x}+a_2k_z\hat{\bm y}+(a_3k_x+a_4k_y)\hat{\bm z}$ \\ \hline
    $\mathbf{D}_2$   & $\mathbf{D}_{2h}$ & $\Gamma_5$  & $a_1k_x\hat{\bm x}+a_2k_y\hat{\bm y}+a_3k_z\hat{\bm z}$  \\ \hline
    $\mathbf{C}_{2v}$  & $\mathbf{D}_{2h}$ & $\Gamma_5$  & $a_1k_y\hat{\bm x}+a_2k_x\hat{\bm y}+a_3k_xk_yk_z\hat{\bm z}$ \\ \hline
    $\mathbf{C}_4$   & $\mathbf{C}_{4h}$ & $(\Gamma_5,\Gamma_6)$    & $(a_1k_x+a_2k_y)\hat{\bm x}+(-a_2k_x+a_1k_y)\hat{\bm y}+a_3k_z\hat{\bm z}$  \\ \hline
    $\mathbf{S}_4$   & $\mathbf{C}_{4h}$ & $(\Gamma_5,\Gamma_6)$    & $(a_1k_x+a_2k_y)\hat{\bm x}+(a_2k_x-a_1k_y)\hat{\bm y}+(bk_+^2+b^*k_-^2)k_z\hat{\bm z}$  \\ \hline
    $\mathbf{D}_4$    & $\mathbf{D}_{4h}$ & $\Gamma_6$  & $a_1(k_x\hat{\bm x}+k_y\hat{\bm y})+a_2k_z\hat{\bm z}$ \\ \hline
    $\mathbf{C}_{4v}$  & $\mathbf{D}_{4h}$ & $\Gamma_6$  & $a_1(k_y\hat{\bm x}-k_x\hat{\bm y})+a_2(k_x^2-k_y^2)k_xk_yk_z\hat{\bm z}$  \\ \hline
    $\mathbf{D}_{2d}$   & $\mathbf{D}_{4h}$ & $\Gamma_6$  & $a_1(k_x\hat{\bm x}-k_y\hat{\bm y})+a_2(k_x^2-k_y^2)k_z\hat{\bm z}$  \\ \hline
    $\mathbf{C}_3$    & $\mathbf{C}_{3i}$ & $(\Gamma_4,\Gamma_5)$    & $(a_1k_x+a_2k_y)\hat{\bm x}+(-a_2k_x+a_1k_y)\hat{\bm y}+a_3k_z\hat{\bm z}$  \\ \hline
    $\mathbf{D}_3$     & $\mathbf{D}_{3d}$ & $\Gamma_4$  & $a_1(k_x\hat{\bm x}+k_y\hat{\bm y})+a_2k_z\hat{\bm z}$  \\ \hline
    $\mathbf{C}_{3v}$  & $\mathbf{D}_{3d}$ & $\Gamma_4$  & $a_1(k_y\hat{\bm x}-k_x\hat{\bm y})+a_2(k_+^3+k_-^3)\hat{\bm z}$ \\ \hline
    $\mathbf{C}_6$    & $\mathbf{C}_{6h}$ & $(\Gamma_7,\Gamma_8)$    & $(a_1k_x+a_2k_y)\hat{\bm x}+(-a_2k_x+a_1k_y)\hat{\bm y}+a_3k_z\hat{\bm z}$ \\ \hline
    $\mathbf{C}_{3h}$  & $\mathbf{C}_{6h}$ & $(\Gamma_7,\Gamma_8)$   & $(b_1k_+^2+b_1^*k_-^2)k_z\hat{\bm x}+i(b_1k_+^2-b_1^*k_-^2)k_z\hat{\bm y}+(b_2k_+^3+b_2^*k_-^3)\hat{\bm z}$ \\ \hline
    $\mathbf{D}_6$  & $\mathbf{D}_{6h}$ & $\Gamma_7$  & $a_1(k_x\hat{\bm x}+k_y\hat{\bm y})+a_2k_z\hat{\bm z}$  \\ \hline
    $\mathbf{C}_{6v}$   & $\mathbf{D}_{6h}$ & $\Gamma_7$  & $a_1(k_y\hat{\bm x}-k_x\hat{\bm y})+ia_2(k_+^6-k_-^6)k_z\hat{\bm z}$  \\ \hline
    $\mathbf{D}_{3h}$   & $\mathbf{D}_{6h}$ & $\Gamma_7$  & $a_1[(k_x^2-k_y^2)k_z\hat{\bm x}-2k_xk_yk_z\hat{\bm y}]+a_2(k_+^3+k_-^3)\hat{\bm z}$  \\ \hline
    $\mathbf{T}$   & $\mathbf{T}_{h}$ & $\Gamma_5$  & $a(k_x\hat{\bm x}+k_y\hat{\bm y}+k_z\hat{\bm z})$  \\ \hline
    $\mathbf{O}$   & $\mathbf{O}_{h}$ & $\Gamma_6$  & $a(k_x\hat{\bm x}+k_y\hat{\bm y}+k_z\hat{\bm z})$  \\ \hline
    $\mathbf{T}_d$  & $\mathbf{O}_{h}$ & $\Gamma_6$  & $a[k_x(k_y^2-k_z^2)\hat{\bm x}+k_y(k_z^2-k_x^2)\hat{\bm y}+k_z(k_x^2-k_y^2)\hat{\bm z}]$  \\ \hline
\end{tabular}
\label{table: gammas-pseudospin}
\end{table}

\subsubsection{Non-pseudospin band}
\label{sec: Rashba-non-pseudospin}

Suppose that the eigenstates of $\hat H_s$ at the $\Gamma$ point in a tetragonal, trigonal, hexagonal, or cubic crystal transform according to a 2D exceptional corep, 
see Table \ref{table: Gamma-point-corep-matrices}. 
Using Eq. (\ref{general-prescription}) one can obtain the ${\cal R}$ matrices for the generators of each point group $\mathbb{G}$. As an example we consider a tetragonal crystal with $\mathbb{G}=\mathbf{D}_{2d}$.
This point group is generated by the roto-reflection $S^-_{4z}=IC^+_{4z}$ and the rotation $C_{2y}$. Since $\mathbb{G}_s=\mathbf{D}_{4h}$ and $\tilde{\mathbb{G}}=\mathbf{D}_4$, there are
two non-pseudospin coreps at the $\Gamma$ point, $\Gamma^+_7$ and $\Gamma^-_7$, see Table \ref{table: coreps-point}. From Eq. (\ref{Gamma-pm}) and Table \ref{table: Gamma-point-corep-matrices} we obtain: 
$\hat{\cal D}_{\Gamma_7^\pm}(S^-_{4z})=\mp\hat D^{(1/2)}(C^+_{4z})$ and $\hat{\cal D}_{\Gamma_7^\pm}(C_{2y})=\hat D^{(1/2)}(C_{2y})$.								
Therefore, $\hat{\cal R}(S^-_{4z})=\hat R(C^+_{4z})$ and $\hat{\cal R}(C_{2y})=\hat R(C_{2y})$.

In a similar fashion, one can show that for all five tetragonal noncentrosymmetric point groups $\mathbf{C}_4$, $\mathbf{S}_4$, $\mathbf{D}_4$, $\mathbf{C}_{4v}$, and $\mathbf{D}_{2d}$, as well as for the cubic groups 
$\mathbf{O}$ and $\mathbf{T}_d$, we have 
\begin{equation}
\label{vector-field-transform}
  \hat{\cal R}(R)=\hat R,\quad \hat{\cal R}(IR)=\hat R,
\end{equation}
where $\hat R$ is the rotation matrix, for both proper and improper symmetry elements. Therefore, in all these cases the point-group constraint is still given by Eq. (\ref{gamma-transform-pseudospin}), 
which means that the antisymmetric SOC transforms as a pseudovector field in the reciprocal space and has the same form as in 
Table \ref{table: gammas-pseudospin}. 

In contrast, in the trigonal and hexagonal systems the symmetry of $\bgam(\bk)$ essentially depends on the $\Gamma$-point corep. For $\mathbb{G}=\mathbf{C}_6$ and $\mathbf{C}_{3h}$, we have 
$\mathbb{G}_s=\mathbf{C}_{6h}$, and, as evident from Table \ref{table: Gamma-point-corep-matrices}, the expression (\ref{vector-field-transform}) holds in the $(\Gamma_9,\Gamma_{10})$ bands. Similarly, 
for $\mathbb{G}=\mathbf{D}_6$, $\mathbf{C}_{6v}$, and $\mathbf{D}_{3h}$, we have $\mathbb{G}_s=\mathbf{D}_{6h}$ and Eq. (\ref{vector-field-transform}) holds in the $\Gamma_8$ bands.
Therefore, the symmetry of $\bgam(\bk)$ in all these non-pseudospin bands is the same as that in the pseudospin ones. In the remaining cases from Table \ref{table: Gamma-point-corep-matrices}, 
$\bgam(\bk)$ does not transform as a pseudovector field under the point group operations. The expressions for the antisymmetric SOC applicable in the vicinity of the $\Gamma$ point are presented 
in Table \ref{table: gammas-nonpseudospin}. In the cases admitting direct comparison, our results agree with Ref. \cite{Smidman-review}.

\begin{table}
\caption{The antisymmetric SOC near the $\Gamma$ point in the non-pseudospin bands; 
$a_i$ and $a$ are real constants, $b_i$ and $b$ are complex constants, and $k_\pm=k_x\pm ik_y$. For each noncentrosymmetric point group $\mathbb{G}$, the corresponding group $\mathbb{G}_s$ and its 2D non-pseudospin 
coreps at the $\Gamma$ point are listed in the second and third columns, respectively.}
\begin{tabular}{|c|c|c|c|}
    \hline
    $\quad \mathbb{G}\quad $ & $\mathbb{G}_s$ & $\Gamma$ & $\bgam(\bk)$ \\ \hline
    $\mathbf{C}_4$   & \ $\mathbf{C}_{4h}$\ \ & $(\Gamma_7,\Gamma_8)$   & $(a_1k_x+a_2k_y)\hat{\bm x}+(-a_2k_x+a_1k_y)\hat{\bm y}+a_3k_z\hat{\bm z}$  \\ \hline
    $\mathbf{S}_4$   & $\mathbf{C}_{4h}$ & $(\Gamma_7,\Gamma_8)$   & $(a_1k_x+a_2k_y)\hat{\bm x}+(a_2k_x-a_1k_y)\hat{\bm y}+(bk_+^2+b^*k_-^2)k_z\hat{\bm z}$  \\ \hline
    $\mathbf{D}_4$    & $\mathbf{D}_{4h}$ & $\Gamma_7$  &  $a_1(k_x\hat{\bm x}+k_y\hat{\bm y})+a_2k_z\hat{\bm z}$ \\ \hline
    $\mathbf{C}_{4v}$  & $\mathbf{D}_{4h}$ & $\Gamma_7$  &  $a_1(k_y\hat{\bm x}-k_x\hat{\bm y})+a_2(k_x^2-k_y^2)k_xk_yk_z\hat{\bm z}$  \\ \hline
    $\mathbf{D}_{2d}$   & $\mathbf{D}_{4h}$ & $\Gamma_7$  & $a_1(k_x\hat{\bm x}-k_y\hat{\bm y})+a_2(k_x^2-k_y^2)k_z\hat{\bm z}$  \\ \hline
    $\mathbf{C}_3$    & $\mathbf{C}_{3i}$ & $\Gamma_6$  & $a_1k_z\hat{\bm x}+a_2k_z\hat{\bm y}+a_3k_z\hat{\bm z}$  \\ \hline
    $\mathbf{D}_3$     & $\mathbf{D}_{3d}$ & $(\Gamma_5,\Gamma_6)$  & $a_1k_z\hat{\bm x}+ia_2(k_+^3-k_-^3)\hat{\bm y}+a_3k_z\hat{\bm z}$  \\ \hline
    $\mathbf{C}_{3v}$  & $\mathbf{D}_{3d}$ & $(\Gamma_5,\Gamma_6)$  & $ia_1(k_+^3-k_-^3)\hat{\bm x}+a_2k_z\hat{\bm y}+ia_3(k_+^3-k_-^3)\hat{\bm z}$ \\ \hline
    $\mathbf{C}_6$    & $\mathbf{C}_{6h}$ & $(\Gamma_9,\Gamma_{10})$  & $(a_1k_x+a_2k_y)\hat{\bm x}+(-a_2k_x+a_1k_y)\hat{\bm y}+a_3k_z\hat{\bm z}$ \\ 
                      &  & $(\Gamma_{11},\Gamma_{12})$  &  $(b_1k_+^3+b_1^*k_-^3)\hat{\bm x}+(b_2k_+^3+b_2^*k_-^3)\hat{\bm y}+ak_z\hat{\bm z}$ \\ \hline
    $\mathbf{C}_{3h}$  & $\mathbf{C}_{6h}$ & $(\Gamma_9,\Gamma_{10})$  & \ $(b_1k_+^2+b_1^*k_-^2)k_z\hat{\bm x}+i(b_1k_+^2-b_1^*k_-^2)k_z\hat{\bm y}+(b_2k_+^3+b_2^*k_-^3)\hat{\bm z}$\ \  \\ 
                      &  & $(\Gamma_{11},\Gamma_{12})$  &  $a_1k_z\hat{\bm x}+a_2k_z\hat{\bm y}+(bk_+^3+b^*k_-^3)\hat{\bm z}$ \\ \hline
    $\mathbf{D}_6$  & $\mathbf{D}_{6h}$ & $\Gamma_8$  & $a_1(k_x\hat{\bm x}+k_y\hat{\bm y})+a_2k_z\hat{\bm z}$ \\ 
                      &   & $\Gamma_9$  &  $a_1(k_+^3+k_-^3)\hat{\bm x}+ia_2(k_+^3-k_-^3)\hat{\bm y}+a_3k_z\hat{\bm z}$  \\ \hline
    $\mathbf{C}_{6v}$   & $\mathbf{D}_{6h}$ & $\Gamma_8$  & $a_1(k_y\hat{\bm x}-k_x\hat{\bm y})+ia_2(k_+^6-k_-^6)k_z\hat{\bm z}$  \\ 
                      &   & $\Gamma_9$  &  $ia_1(k_+^3-k_-^3)\hat{\bm x}+a_2(k_+^3+k_-^3)\hat{\bm y}+ia_3(k_+^6-k_-^6)k_z\hat{\bm z}$ \\ \hline
    $\mathbf{D}_{3h}$   & $\mathbf{D}_{6h}$ & $\Gamma_8$  & $a_1[(k_x^2-k_y^2)k_z\hat{\bm x}-2k_xk_yk_z\hat{\bm y}]+a_2(k_+^3+k_-^3)\hat{\bm z}$  \\ 
                      &   & $\Gamma_9$  &  $a_1k_z\hat{\bm x}+ia_2(k_+^6-k_-^6)k_z\hat{\bm y}+a_3(k_+^3+k_-^3)\hat{\bm z}$   \\ \hline
    $\mathbf{O}$   & $\mathbf{O}_{h}$ & $\Gamma_7$  & $a(k_x\hat{\bm x}+k_y\hat{\bm y}+k_z\hat{\bm z})$  \\ \hline
    $\mathbf{T}_d$  & $\mathbf{O}_{h}$ & $\Gamma_7$  & $a[k_x(k_y^2-k_z^2)\hat{\bm x}+k_y(k_z^2-k_x^2)\hat{\bm y}+k_z(k_x^2-k_y^2)\hat{\bm z}]$  \\ \hline
\end{tabular}
\label{table: gammas-nonpseudospin}
\end{table}

To illustrate our procedure, let us consider the case of the $\Gamma_9$ bands for $\mathbb{G}=\mathbf{D}_{3h}$, which is generated by the roto-reflection $S^-_{3z}=IC^+_{6z}$ and the rotation $C_{2y}$. We have
$\mathbb{G}_s=\mathbf{D}_{6h}$ and $\tilde{\mathbb{G}}=\mathbf{D}_6$. For the non-pseudospin coreps $\Gamma^\pm_9$ of $\mathbf{D}_{6h}$, see Table \ref{table: Gamma-point-corep-matrices}, we have
$\hat{\cal D}_{\Gamma_9^\pm}(S^-_{3z})=\pm i\hat\sigma_3$ and $\hat{\cal D}_{\Gamma_9^\pm}(C_{2y})=\hat D^{(1/2)}(C_{2y})$. Therefore, $\hat{\cal R}(S^-_{3z})=\hat R(C_{2z})$ and $\hat{\cal R}(C_{2y})=\hat R(C_{2y})$.								
From Eq. (\ref{gamma-transform-gen}), the symmetry constraints take the following form: 
\begin{eqnarray*}
  && \gamma_{x,y}(k_+,k_-,k_z)=\gamma_{x,y}(e^{-i\pi/3}k_+,e^{i\pi/3}k_-,k_z),\ \gamma_z(k_+,k_-,k_z)=-\gamma_z(e^{-i\pi/3}k_+,e^{i\pi/3}k_-,k_z),\\
  && \gamma_{x,z}(k_+,k_-,k_z)=\gamma_{x,z}(k_-,k_+,k_z),\ \gamma_y(k_+,k_-,k_z)=-\gamma_y(k_-,k_+,k_z)
\end{eqnarray*}
where $k_\pm=k_x\pm ik_y$. The lowest-order odd degree polynomial solutions of these equations are given by $\gamma_x\propto k_z$, $\gamma_y\propto(k_+^6-k_-^6)k_z$, and $\gamma_z\propto k_+^3+k_-^3$.

\subsection{Two-band Rashba model}
\label{sec: two-band Rashba}

The symmetry analysis of the previous subsection can be extended to the multiband case. The possibility that the states $|\bk,n,s\rangle$ in different bands transform according to different coreps 
can be accounted for by introducing an additional band index in Eq. (\ref{general-prescription}):
\begin{equation}
\label{general-Bloch-transform-multiband}
    g|\bk,n,s\rangle=\sum_{s'}|g\bk,s'\rangle {\cal D}_{n,s's}(g). 
\end{equation}
Here $\hat{\cal D}_n(g)$ is the $\Gamma$-point corep in the $n$th band. The matrix elements of the antisymmetric part of the Hamiltonian are given by Eq. (\ref{H_a-matrix-elements}) and we obtain:
\begin{eqnarray*}
  \langle\bk,n,s|g^\dagger \hat H_a g|\bk,n',s'\rangle=\sum_{s_1s_2}{\cal D}_{n,s_1s}^*(g){\cal D}_{n',s_2s'}(g)\langle g\bk,n,s_1|\hat H_a|g\bk,n',s_2\rangle \\
  =iA_{nn'}(g\bk)[\hat{\cal D}_n^\dagger(g)\hat{\cal D}_{n'}(g)]_{ss'}+\bB_{nn'}(g\bk)[\hat{\cal D}_n^\dagger(g)\hat{\bm{\sigma}}\hat{\cal D}_{n'}(g)]_{ss'},
\end{eqnarray*}
instead of Eq. (\ref{H_a-matrix-transform}). Since $\hat H_a$ commutes with all $g\in\mathbb{G}$, the symmetry constraint on the parameters $A$ and $\bm{B}$ takes the 
following form:
\begin{equation}
\label{AB-symmetry-constraint}
  iA_{nn'}(\bk)\hat\sigma_0+\bB_{nn'}(\bk)\hat{\bm{\sigma}}=
  iA_{nn'}(g\bk)[\hat{\cal D}_n^\dagger(g)\hat{\cal D}_{n'}(g)]+\bB_{nn'}(g\bk)[\hat{\cal D}_n^\dagger(g)\hat{\bm{\sigma}}\hat{\cal D}_{n'}(g)],
\end{equation}
which can be evaluated for each pair of bands. In particular, if the bands $n$ and $n'$ correspond to the same corep, $\hat{\cal D}_{n}(g)=\hat{\cal D}_{n'}(g)=\hat{\cal D}(g)$, we have
\begin{equation}
\label{AB-same-corep}
  A_{nn'}(\bk)=A_{nn'}(g^{-1}\bk),\quad B_{nn',i}(\bk)=\sum_{j=1}^3{\cal R}_{ij}(g)B_{nn',j}(g^{-1}\bk),\quad g\in\mathbb{G},
\end{equation}
where the ${\cal R}$ matrix is defined in Eq. (\ref{sigma-rotation-non-pseudospin}). If both bands are pseudospin bands, then $A$ transforms as an invariant scalar field, while $\bB$ transforms as 
an invariant vector field, see Ref. \cite{Sam09}. However, if one of the bands is not a pseudospin band, then the symmetry properties of the effective SO Hamiltonian become more complicated.

Due to a large number of possibilities, here we consider only the case of two bands in a tetragonal crystal with $\mathbb{G}=\mathbf{C}_{4v}$. This point group describes, for
instance, the symmetry of CePt$_3$Si and other popular noncentrosymmetric systems \cite{NCSC-book}. Introducing the notations
\begin{eqnarray*}
  && A_{11}=A_{22}=0,\quad A_{12}=-A_{21}=\alpha,\\
  && \bm{B}_{11}=\bgam_1,\quad \bm{B}_{22}=\bgam_2,\quad \bm{B}_{12}=\bm{B}_{21}=\tilde{\bgam},
\end{eqnarray*}
the Hamiltonian (\ref{H-second-quant}) takes the form of two coupled Rashba models: 
\begin{equation}
\label{H-4-band}
  \hat H=\hat H_1+\hat H_2+\hat H_{12},
\end{equation}
where
$$
  \hat H_n=\sum\limits_{\bk,ss'}[\epsilon_n(\bk)\delta_{ss'}+\bgam_n(\bk)\bm{\sigma}_{ss'}]\hat a^\dagger_{\bk ns}\hat a_{\bk ns'}
$$
and 
$$
  \hat H_{12}=\sum\limits_{\bk,ss'}[i\alpha(\bk)\delta_{ss'}+\tilde{\bgam}(\bk)\bm{\sigma}_{ss'}]\hat a^\dagger_{\bk 1s}\hat a_{\bk 2s'}+\mathrm{H.c.}
$$
Here $\bgam_1$ and $\bgam_2$ are real and odd in $\bk$, while $\alpha$ and $\tilde{\bgam}$ are real and odd (even) in $\bk$, if the bands have the same (opposite) parity. 
Since the intraband Rashba couplings have been studied in Sec. \ref{sec: one-band Rashba}, see Tables \ref{table: gammas-pseudospin} and
\ref{table: gammas-nonpseudospin}, below we focus only on the properties of $\alpha(\bk)$ and $\tilde{\bgam}(\bk)$.

According to Table \ref{table: coreps-point}, the group $\mathbb{G}_s=\mathbf{D}_{4h}$ has four double-valued coreps, $\Gamma_6^\pm$ and $\Gamma_7^\pm$ (only $\Gamma_6^+$ is a pseudospin one), 
which leads to ten possible two-band combinations: $n=\Gamma_6^+$, $n'=\Gamma_6^+$, \textit{etc}. Using Eq. (\ref{Gamma-pm}) and Table \ref{table: Gamma-point-corep-matrices}, 
we obtain the subduced corep matrices for the generators of $\mathbb{G}=\mathbf{C}_{4v}$:
$$
  \hat{\cal D}_{\Gamma_6^p}(C^+_{4z})=\hat D^{(1/2)}(C^+_{4z}),\quad 
  \hat{\cal D}_{\Gamma_6^p}(\sigma_y)=p\hat D^{(1/2)}(C_{2y})
$$
and
$$
\label{corep-g-Gamma_7}
  \hat{\cal D}_{\Gamma_7^p}(C^+_{4z})=-\hat D^{(1/2)}(C^+_{4z}),\quad 
  \hat{\cal D}_{\Gamma_7^p}(\sigma_y)=p\hat D^{(1/2)}(C_{2y}),
$$
where $p=\pm$ is the parity index. For the $(\Gamma_6^p,\Gamma_6^{p'})$ and $(\Gamma_7^p,\Gamma_7^{p'})$ pairs of bands, the substitution of the above matrices in Eq. (\ref{AB-symmetry-constraint}) 
produces the following symmetry-imposed constraints:
$$
  \alpha(\bk)=\alpha(C^-_{4z}\bk),\ \alpha(\bk)=pp'\alpha(\sigma_y\bk),\ 
  \tilde\bgam(\bk)=C^+_{4z}\tilde\bgam(C^-_{4z}\bk),\ \tilde\bgam(\bk)=pp'C_{2y}\tilde\bgam(\sigma_y\bk).
$$
In contrast, in the case of $(\Gamma_6^p,\Gamma_7^{p'})$ bands we obtain:
$$
  \alpha(\bk)=-\alpha(C^-_{4z}\bk),\ \alpha(\bk)=pp'\alpha(\sigma_y\bk),\ 
  \tilde\bgam(\bk)=-C^+_{4z}\tilde\bgam(C^-_{4z}\bk),\ \tilde\bgam(\bk)=pp'C_{2y}\tilde\bgam(\sigma_y\bk).
$$
Representative expressions for even and odd $\alpha$ and $\tilde{\bgam}$ satisfying these constraints are given in Table \ref{table: two-band Rashba}. 
One can see that, unlike $\bgam_1$ and $\bgam_2$, the interband couplings are sensitive to the relative parity of the bands.

\begin{table}
\caption{The interband couplings of the two-band Rashba model near the $\Gamma$ point, for $\mathbb{G}=\mathbf{C}_{4v}$; $a_{0,1,2}$ are real constants.}
\begin{tabular}{|c|c|c|}
    \hline
    $\quad n,n'\quad $  &  $\alpha(\bk)$ & $\tilde{\bgam}(\bk)$  \\ \hline
    $\Gamma_6^+,\Gamma_6^+$   &  $a_0k_z$ &  \ $a_1(k_y\hat{\bm x}-k_x\hat{\bm y})+a_2(k_x^2-k_y^2)k_xk_yk_z\hat{\bm z}$\ \  \\ \hline
    $\Gamma_6^+,\Gamma_6^-$   &  \ $a_0(k_x^2-k_y^2)k_xk_y$\ \   &  $a_1(k_xk_z\hat{\bm x}+k_yk_z\hat{\bm y})+a_2\hat{\bm z}$ \\ \hline
    $\Gamma_6^-,\Gamma_6^-$   &  $a_0k_z$  &  $a_1(k_y\hat{\bm x}-k_x\hat{\bm y})+a_2(k_x^2-k_y^2)k_xk_yk_z\hat{\bm z}$ \\ \hline	                          
    $\Gamma_7^+,\Gamma_7^+$   &  $a_0k_z$  &  $a_1(k_y\hat{\bm x}-k_x\hat{\bm y})+a_2(k_x^2-k_y^2)k_xk_yk_z\hat{\bm z}$ \\ \hline
    $\Gamma_7^+,\Gamma_7^-$   &  $a_0(k_x^2-k_y^2)k_xk_y$  &  $a_1(k_xk_z\hat{\bm x}+k_yk_z\hat{\bm y})+a_2\hat{\bm z}$ \\ \hline
    $\Gamma_7^-,\Gamma_7^-$   &  $a_0k_z$  &  $a_1(k_y\hat{\bm x}-k_x\hat{\bm y})+a_2(k_x^2-k_y^2)k_xk_yk_z\hat{\bm z}$ \\ \hline                          
    $\Gamma_6^+,\Gamma_7^+$   &  $a_0(k_x^2-k_y^2)k_z$  &  $a_1(k_y\hat{\bm x}+k_x\hat{\bm y})+a_2k_xk_yk_z\hat{\bm z}$ \\ \hline
    $\Gamma_6^+,\Gamma_7^-$   &  $a_0k_xk_y$  &  $a_1(k_xk_z\hat{\bm x}-k_yk_z\hat{\bm y})+a_2(k_x^2-k_y^2)\hat{\bm z}$ \\ \hline
    $\Gamma_6^-,\Gamma_7^+$   &  $a_0k_xk_y$  &  $a_1(k_xk_z\hat{\bm x}-k_yk_z\hat{\bm y})+a_2(k_x^2-k_y^2)\hat{\bm z}$ \\ \hline
    $\Gamma_6^-,\Gamma_7^-$   &  $a_0(k_x^2-k_y^2)k_z$  &  $a_1(k_y\hat{\bm x}+k_x\hat{\bm y})+a_2k_xk_yk_z\hat{\bm z}$ \\ \hline                        
\end{tabular}
\label{table: two-band Rashba}
\end{table}

\subsection{Band degeneracies}
\label{sec: band degeneracies}

The spectrum of the Hamiltonian (\ref{H-4-band}) consists of four bands $\xi_{1,2,3,4}$, which can be obtained by diagonalizing the following $4\times 4$ matrix:
\begin{equation}
\label{4-band-epsilon}
  \hat\varepsilon(\bk)=\left( \begin{array}{cc}
                             \epsilon_1(\bk)+\bgam_1(\bk)\hat{\bm{\sigma}} & i\alpha(\bk)+\tilde{\bgam}(\bk)\hat{\bm{\sigma}} \\
                             -i\alpha(\bk)+\tilde{\bgam}(\bk)\hat{\bm{\sigma}} & \epsilon_2(\bk)+\bgam_2(\bk)\hat{\bm{\sigma}}
                             \end{array} \right).
\end{equation}
The bands are completely split at almost all $\bk$, except some high symmetry locations.  
Using Tables \ref{table: gammas-pseudospin}, \ref{table: gammas-nonpseudospin}, and \ref{table: two-band Rashba}, it is easy to see that the bands remain twofold degenerate along the whole $\Lambda$ line. 
Indeed, for all combinations of the bands, the intraband Rashba couplings $\bgam_{1,2}$ vanish at $k_x=k_y=0$. Then, the eigenvalues of the matrix (\ref{4-band-epsilon}) come in pairs given by 
$$
  \xi_{1,2}=\frac{\epsilon_1+\epsilon_2}{2}+\sqrt{\left(\frac{\epsilon_1-\epsilon_2}{2}\right)^2+\alpha^2+|\tilde{\bgam}|^2},\quad
  \xi_{3,4}=\frac{\epsilon_1+\epsilon_2}{2}-\sqrt{\left(\frac{\epsilon_1-\epsilon_2}{2}\right)^2+\alpha^2+|\tilde{\bgam}|^2},
$$
for any values of the interband parameters $\alpha$ and $\tilde{\bgam}$.

The inevitable twofold degeneracy of the bands along the $\Lambda$ line and the band splitting at all other $\bk$ can be understood using a simple symmetry argument. 
In a noncentrosymmetric crystal, the lattice potential is no longer invariant under inversion $I$ and the full symmetry group of the reduced Hamiltonian (\ref{H_k}) at $\bk\neq\bm{0}$ is given by 
\begin{equation}
\label{G_k-NC}
  {\cal G}_{\bk}=G_{\bk},
\end{equation}
instead of Eq. (\ref{magnetic G_k}). This group does not contain any antiunitary elements, neither $K$ nor ${\cal C}$, and is just a nonmagnetic point group. 
Therefore, the eigenstates of $\hat H_{\bk}$ can be classified according to the usual double-valued irreps of $G_{\bk}$, instead of coreps. 

In the case of $\mathbb{G}=\mathbf{C}_{4v}$, we have $G_{\bk}=\mathbf{C}_{4v}$ along the $\Lambda$ line. Since both double-valued irreps of this group, $\Gamma_6$ and $\Gamma_7$, are 2D, 
see Table \ref{table: coreps-lines}, the bands have to be twofold degenerate along the $\Lambda$ line. In contrast, for a general $\bk$ we have $G_{\bk}=\mathbf{C}_1$, which has just one double-valued irrep $\Gamma_2$, 
see Sec. \ref{sec: general k}. Since this irrep is 1D, the bands are nondegenerate at a general $\bk$. If $\bk$ is in a high symmetry plane, then $G_{\bk}=\mathbf{C}_s$. This group has two double-valued irreps, 
$\Gamma_3$ and $\Gamma_4$, both 1D, see Sec. \ref{sec: planes}. In the absence of an additional antiunitary symmetry of $\hat H_{\bk}$, these two complex conjugate irreps remain nondegenerate, 
thus lifting the band degeneracy in the special planes.

\section{Conclusions}
\label{sec: Conclusion}

The electron Bloch bands in centrosymmetric crystals are at least twofold degenerate at all wave vectors $\bk$ and can be classified according to the irreducible coreps of the magnetic group of $\bk$. 
Since these coreps are not necessarily equivalent to the spin-$1/2$ corep, the Bloch states do not always transform under the point group operations and time reversal in the same way as the pure spin-$1/2$ eigenstates and, 
therefore, cannot be characterized by a pseudospin quantum number, in general. While the inversion-even bands in triclinic, monoclinic, and orthorhombic crystals are all pseudospin bands, 
the pseudospin description fails for the inversion-odd bands in all crystal systems and also for some inversion-even bands in tetragonal, trigonal, hexagonal, and cubic crystals. 

We propose a generalization of the Ueda-Rice formula to define the relative orientations of the Bloch bases at different $\bk$ in any twofold degenerate band, pseudospin or non-pseudospin. 
This prescription, see Eqs. (\ref{general-prescription}), (\ref{general-prescription-I}), and (\ref{general-prescription-K}), is compatible with all local transformation properties of the Bloch states at the 
high symmetry locations in the BZ. As an application of the formalism, we derive the effective Hamiltonians of the electron-lattice SOC in crystals without an inversion center. The complete classification of the
single-band Rashba Hamiltonians for all noncentrosymmetric point groups and the two-band Rashba Hamiltonians in the tetragonal case is presented.  
We have shown that the expressions for the intraband and interband Rashba couplings involving non-pseudospin bands are considerably different from the ones obtained previously for pseudospin bands.

\acknowledgments

This work was supported by a Discovery Grant 2015-06656 from the Natural Sciences and Engineering Research Council of Canada.

\appendix

\section{Corepresentations of magnetic point groups}
\label{app: coreps}

In this Appendix, we summarize the relevant properties of the coreps of magnetic groups. The detailed explanations and proofs can be found in Refs. \cite{BC-book} and \cite{BD68}. 
Suppose we have a magnetic group ${\cal G}=G+AG$, where the group $G$ is the unitary component and $A$ is an antiunitary operation. 
In our case, $G$ is the double group of $\bk$ and $A$ is the conjugation operation ${\cal C}$, see Eqs. (\ref{magnetic G_k}) and (\ref{magnetic G_0}),
so that $A$ commutes with all elements of $G$ and $A^2=-1$ when acting on spin-1/2 wave functions. 

Let $\Gamma$ be an irrep of $G$ of dimension $d$, with the basis functions $\phi_1,...,\phi_d$, such that for any element $g\in G$ we have
\begin{equation}
\label{app: irrep-transform}
  g\phi_i=\sum_{j=1}^d\phi_jD_{ji}(g),
\end{equation}
where $\hat D$ is the unitary representation matrix. We also introduce another $d$ functions $\bar\phi_1,...,\bar\phi_d$, such that $\bar\phi_i=A\phi_i$. Then, the action of the unitary ($g\in G$)
and antiunitary ($a\in AG$) elements of ${\cal G}$ on the $2d$ functions $(\phi,\bar\phi)$ is given by $g(\phi,\bar\phi)=(\phi,\bar\phi)\hat {\cal D}(g)$ and $a(\phi,\bar\phi)=(\phi,\bar\phi)\hat {\cal D}(a)$, 
respectively. Here
\begin{equation}
\label{app: corep-matrices}
  \hat{\cal D}(g)=\left(\begin{array}{cc}
              \hat D(g) & 0 \\
              0 & \hat D^*(g)
              \end{array}\right),\qquad 
  \hat{\cal D}(a)=\left(\begin{array}{cc}
              0 & \hat D(aA) \\
              \hat D^*(A^{-1}a) & 0
              \end{array}\right)
\end{equation}
are $2d\times 2d$ unitary matrices forming the corepresentation of ${\cal G}$ derived from the irrep $\Gamma$. 

The multiplication rules for the corepresentation matrices are different from those for the usual irreps. For $g,g_1,g_2\in G$ and $a,a_1,a_2\in AG$ we have
$\hat{\cal D}(g_1g_2)=\hat{\cal D}(g_1)\hat{\cal D}(g_2)$, $\hat{\cal D}(ga)=\hat{\cal D}(g)\hat{\cal D}(a)$, $\hat{\cal D}(ag)=\hat{\cal D}(a)\hat{\cal D}^*(g)$, 
and $\hat{\cal D}(a_1a_2)=\hat{\cal D}(a_1)\hat{\cal D}^*(a_2)$. The presence of the complex conjugate matrices here and in Eqs. (\ref{app: corep-matrices}) reflects the antilinearity of $A$. 
We note that, since any antiunitary element can be written in the form $a=Ag$, where $g\in G$, we have
$$
  \hat{\cal D}(Ag)=\left(\begin{array}{cc}
              0 & -\hat D(g) \\
              \hat D^*(g) & 0
              \end{array}\right),
$$
since $A^2=-1$. Therefore, the corepresentation of all antiunitary elements can be obtained from 
\begin{equation}
\label{app: A matrix}
  \hat{\cal D}(A)=\left(\begin{array}{cc}
              0 & -\mathbb{1}_d \\
              \mathbb{1}_d & 0
              \end{array}\right),
\end{equation}
where $\mathbb{1}_d$ is $d\times d$ unit matrix, by applying the multiplication rules given above.

Similarly to the usual group representations, the corepresentation is said to be reducible if the matrices (\ref{app: corep-matrices}) can be brought to a block-diagonal form by a unitary transformation. 
Whether $\hat{\cal D}$ is reducible or not depends on the relation between the irreps $\hat D$ and $\hat D^*$. There are three possibilities, called Case A, Case B, and Case C.  
If $\hat D$ and $\hat D^*$ are equivalent, then there exists a unitary matrix $\hat V$ such that $\hat D(g)=\hat V\hat D^*(g)\hat V^{-1}$. One can show that $\hat V\hat V^*=\pm\hat D(A^2)=\mp\mathbb{1}_d$. 
If the upper sign is realized (Case A), then the corep $\hat{\cal D}$ is reducible, while for the lower sign (Case B) the corep $\hat{\cal D}$ is irreducible. 
If $\hat D$ and $\hat D^*$ are inequivalent, then the corep $\hat{\cal D}$ is irreducible (Case C). In the literature, different names for these three cases are sometimes used: Case A is ``pseudoreal'' or Type 2,
Case B is ``real'' or Type 1, and Case C is ``complex'' or Type 3, see Ref. \cite{Lax-book}.

The difference between the three cases can be understood as follows. Denoting the $d$-dimensional vector spaces 
spanned by the $\phi$'s and $\bar\phi$'s by $L$ and $\bar L$, respectively, one can show that $\bar L$ is either identical to $L$ or orthogonal to $L$. The Case A corresponds to the former possibility, \textit{i.e.},  the 
set of the conjugate basis functions $\bar\phi_1,...,\bar\phi_d$ is the same as $\phi_1,...,\phi_d$. Therefore, $\dim\hat{\cal D}=\dim\hat D=d$ and for the corepresentation matrices we obtain:
\begin{equation}
\label{app: corep-A}
  \hat{\cal D}(g)=\hat D(g),\qquad \hat{\cal D}(A)=\hat V
\end{equation}
In this case, the presence of the antiunitary symmetry $A$ does not lead to an additional degeneracy. Note that it follows from $\hat V\hat V^*=-\mathbb{1}_d$ that $\hat V$ is an 
antisymmetric unitary matrix, therefore Case A can only be realized if $d$ is even. 

In Case B, $\bar L$ is orthogonal to $L$, $\dim\hat{\cal D}=2d$, and the corepresentation matrices can be brought to the form
\begin{equation}
\label{app: corep-B}
  \hat{\cal D}(g)=\left(\begin{array}{cc}
              \hat D(g) & 0 \\
              0 & \hat D(g)
              \end{array}\right),\qquad 
  \hat{\cal D}(A)=\left(\begin{array}{cc}
              0 & -\hat V \\
              \hat V & 0
              \end{array}\right).
\end{equation}
The additional degeneracy due to the antiunitary symmetry $A$ is said to be of the ``doubling'' type. 

In Case C, $\bar L$ is orthogonal to $L$ and the irreps $\hat D$ and $\hat D^*$ pair up to form a single irreducible corep $\hat{\cal D}$ of twice the dimension, $\dim\hat{\cal D}=2d$. 
The corepresentation matrices have the form 
\begin{equation}
\label{app: corep-C}
  \hat{\cal D}(g)=\left(\begin{array}{cc}
              \hat D(g) & 0 \\
              0 & \hat D^*(g)
              \end{array}\right),\qquad 
  \hat{\cal D}(A)=\left(\begin{array}{cc}
              0 & -\mathbb{1}_d \\
              \mathbb{1}_d & 0
              \end{array}\right)
\end{equation}
In this case, the antiunitary symmetry $A$ leads to an additional degeneracy of the ``pairing'' type. 

There is a quick practical way to determine which of the three corep types is realized for a given irrep $\Gamma$, called the Dimmock-Wheeler test \cite{BD68}, see also Ref. \cite{Herring37}. 
For $A^2=-1$ this test takes the following form:
\begin{equation}
\label{app: Dimmock-Wheeler}
  \sum_{g\in G}\chi_\Gamma(g^2)=\left\{\begin{array}{rl}
                                      -|G| & \ \mathrm{in\ Case\ A},\\
                                      |G| & \ \mathrm{in\ Case\ B},\\
                                      0 & \ \mathrm{in\ Case\ C},
                                      \end{array}\right.
\end{equation}
where the summation goes over all elements of the unitary component $G$, $\chi_\Gamma$ is the character of $\Gamma$, and $|G|$ is the order of $G$.

\subsection{2D coreps of ${\cal G}_{\bk}$}
\label{app: 2D coreps}

Let us take $G=G_{\bk}$ and $A={\cal C}$. According to Refs. \cite{Lax-book,BC-book}, for all double-valued irreps of $G_{\bk}$ at $\bk\neq\bm{0}$ we have either $d=1$, producing 2D Case B or Case C coreps, or $d=2$, 
producing 2D Case A coreps. In Cases B and C, the antiunitary symmetry results 
in twofold degeneracy of the orthogonal conjugate states $\phi$ and $\bar\phi={\cal C}\phi$, where $g\phi=\chi_\Gamma(g)\phi$ and $g\bar\phi=\chi^*_\Gamma(g)\bar\phi$. 
In Case B, the characters are real and Eq. (\ref{app: corep-B}) takes the form
\begin{equation}
\label{app: corep-B-1D}
  \hat {\cal D}_\Gamma(g)=\left(\begin{array}{cc}
              \chi_\Gamma(g) & 0 \\
              0 & \chi_\Gamma(g)
              \end{array}\right),\qquad 
  \hat {\cal D}_\Gamma({\cal C})=\left(\begin{array}{cc}
              0 & -1 \\
              1 & 0
              \end{array}\right).
\end{equation}
In Case C, the characters are complex and Eq. (\ref{app: corep-C}) takes the form
\begin{equation}
\label{app: corep-C-1D}
  \hat {\cal D}_\Gamma(g)=\left(\begin{array}{cc}
              \chi_\Gamma(g) & 0 \\
              0 & \chi^*_\Gamma(g)
              \end{array}\right),\qquad 
  \hat {\cal D}_\Gamma({\cal C})=\left(\begin{array}{cc}
              0 & -1 \\
              1 & 0
              \end{array}\right).
\end{equation}
In Case A, the states $\phi$ and $\bar\phi$ form the basis of a 2D irrep. The matrix $\hat V$ in Eq. (\ref{app: corep-A}) is a $2\times 2$ antisymmetric unitary matrix, which can be chosen in the form 
$\hat V=-i\hat\sigma_2$. Thus we have
\begin{equation}
\label{app: corep-A-2D}
  \hat {\cal D}_\Gamma(g)=\hat D_\Gamma(g),\qquad 
  \hat {\cal D}_\Gamma({\cal C})=\left(\begin{array}{cc}
              0 & -1 \\
              1 & 0
              \end{array}\right).
\end{equation}
One can use Eqs. (\ref{app: corep-B-1D}), (\ref{app: corep-C-1D}), and (\ref{app: corep-A-2D}) for the double-valued coreps at $\bk=\bm{0}$ as well, except the two 4D coreps in cubic crystals mentioned in 
Sec. \ref{sec: Gamma point}.

\end{document}